# Tunable electronic properties of germanene and two-dimensional group-III phosphides heterobilayers

By

Md. Rayid Hasan Mojumder
Roll: 1503066

A thesis dissertation submitted in requirements for the completion of the degree of Bachelor of Science in Electrical and Electronic Engineering from the Department of Electrical and Electronic Engineering

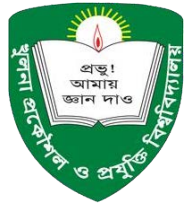

Khulna University of Engineering & Technology
Khulna 9203, Bangladesh

**February 2020**

# DECLARATION

This is to certify that the thesis work entitled **"Study on tunable electronic properties of germanene and two-dimensional group-III phosphides heterobilayers."** has been carried out by ***Md. Rayid Hasan Mojumder*** in the Department of Electrical and Electronic Engineering, Khulna University of Engineering & Technology, Khulna-9203, Bangladesh. The above thesis work or any part of this work has not been submitted anywhere to award any degree or diploma. And the research work is solely conducted without plagiarism or any violation of research ethics and precepts.

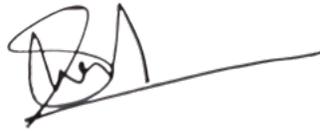

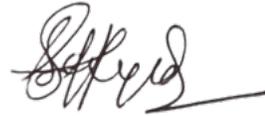

Signature of Supervisor                Signature of Candidate

# ACKNOWLEDGMENT

I would like to express my cordial tribute towards my supervisor and instructor, Dr. Md. Sherajul Islam, Professor, Department of Electrical and Electronic Engineering, Khulna University of Engineering & Technology, Khulna, for his continuous instructions, validations, and inspirations all through the research work and for picking me up with his sublime diligent knowledge on 2D materials. His co-operation and resource materials helped me all the way through this thesis.

I am elated to thank all my teammates and co-workers who have helped me in and out of this research work and always supported me in times of theoretical complexities and simulation difficulties. And I am owed to all the senior researchers under my supervisor for their guidelines to move towards the right direction with the thesis and to bind this very dissertation.

Finally, I would wrap up by showing gratitude towards all the honorable teachers and staff for their direct and indirect supports that helped me to perform the thesis work. My all regards to the research authors and publications who have been carried out their research and duties for a better tomorrow, for a materials revolution.

# ABSTRACT

In this research work, the 2D structure of the germanene layer is compounded with 2D group-III phosphides: AlP and GaP. The planar structure of AlP and low-buckled GaP have been taken to form the bilayer patterns. In each case, three stacking patterns are considered, and their relaxed interlayer distance and binding energy have been reported. The binding energy being around in the range between ~150 to 210 meV shows the existence of weak van der Waals interactions between the layers. The heterostructures containing germanene and these two phosphides show an opening of a large indirect bandgap of magnitude range of ~200 meV to 600 meV, which can be tuned by changing interlayer distance and by incorporating bi-axial compressive and tensile strain. Although their normal bandgap, which significantly changes with SOC, is an indirect one, whilst tunning the interlayer distance band gap jumps from unsymmetrical point to symmetrical Dirac cones and becomes direct on K points. The charge carrier mostly concentrates on the p-orbitals of the germanene in the conduction regions; thus, the electrical properties of germanene will be retained, and the carrier will provide a much faster device response property. The absence of the phosphides influence makes them the intended substrate for growing the germanene layer on top of that. Again, due to the bandgap at Dirac cones being opened and jumps between the Dirac cones and band gap changes with SOC tropological insulator can be formed, and Quantum Spin Hall effect may exist.

# CONTENTS

© Md. Rayid Hasan Mojumder

# LIST OF TABLES

# LIST OF FIGURES

© Md. Rayid Hasan Mojumder

© Md. Rayid Hasan Mojumder

# LIST OF ACRONYMS

| | |
|---|---|
| *2D* | Two Dimensional |
| *h-BN* | Hexagonal Boron Nitride |
| *TMD* | Transition Metal Chalcogenides |
| *QSH* | Quantum Spin Hall |
| *SOC* | Spin Orbit Coupling |
| *Sn* | Stanene |
| *SiC* | Silicon Carbide |
| *Pb* | Plumbene |
| *DFT* | Density Functional Theory |
| *HF* | Hartree-Fock |
| *PDOS* | Partial Density of States |
| *LDA* | Local Density Approximation |
| *QE* | QUANTUM ESPRESSO |
| *PP* | Pseudo-potentials |
| *GGA* | Generalized Gradient Approximation |
| *SCF* | Self-Consistent Field |
| *PWSCF* | Plane Wave Self-Consistent Field |
| *MP* | Monkhorst-Pack |
| *vdW* | Vander Waals |

# CHAPTER I

## Introduction

## 1.1 Introduction

Two-dimensional materials have been a major aspect towards the achievement of prospective characteristics and behavior required for awaiting future devices. The exfoliation of two-dimensional (2D) graphene from graphite in the year 2004 [1] introduced the sublime properties of the 2D materials that would result in a step on the device performance improvements afterward. A wide range of mechanical, thermal, and electrical properties is often associated with pure graphene that is readily synthesized. This enables graphene to become able to improve the performance of various products and materials. The 'enabling' characteristics of graphene exacerbate the use of graphene in new applications as well as replace existing materials that have been used in the manufacture of devices and products; after the invention of graphene, remarkable development has been made in the past decade in the study of 2D materials at the point of condensed matter physics, material science and chemistry. The researcher tends to investigate graphene-like materials and end up having materials: silicene [2,3], germanene [4,5], stanene [6,7], plumbene [8,9], bismuthene [10], hexagonal boron nitride (h-BN) [11], 2D silicon carbide [12], transition metal chalcogenides (TMD) [13], etc. which became some 2D materials of interest [14,15] in the ongoing research on the growing technological nodes. Ultralow weight, high Young's modulus, high strength, outstanding carrier mobility, as well as high anisotropy between the in-plane and out-of-plane mechanical properties all result due to stacking configuration on two dimensions. Again, removal of van der Waals interactions, an increase in the ratio of surface-area-to-volume, confinement of electrons in a plane made them promising for applications as transistors and sensors, photodetectors, battery electrodes, topological insulators, valleytronics, etc. Atoms in the layer of the 2D materials are connected via a covalent bond where weak van der Waals interaction may exist between the interlayer atoms. In addition to the electronic properties, 2D materials have exceptional biochemical optical sensing properties. First of all, atomic-thin layer structure and large surface area make them extraordinary substrates that contain π–π stacking in order to absorb biomolecules. Next, to mention that due to its wide surface-to-volume ratio, 2D material provides high-energy transfer efficacy in addition to the fast response time resulting in ultrafast carrier mobility. 2D materials being highly compatible with sensor technology in the recent future will be available in markets such as wearable electronics, optoelectronics, and semiconductor technology. Figure 1.1 depicts various applications of 2D materials.

These 2D layers can be applied as a monolayer (lateral 2D heterostructure) or a multilayer stack (vertical 2D heterostructure). Research has been conducted in 2D electronics provides the way to use 2D materials as the tunnel MOSFET and in order to fabricate high-grade devices to reduce the contact resistance [16]. In a heterostructure minimum, two different single-layered materials are stacked in a specifically selected sequence is an effective approach to tune the electronic properties [17] and to improve the performance, stability, and quality of the materials [18,19]. This stacking shows new properties and applications that emerged from the layers. Nowadays, group-IV materials like germanene, silicene, plumbene, and stanene, etc., have shown their worth as being applicable in nanodevices owing to their sublime electronic, optical, and mechanical properties.

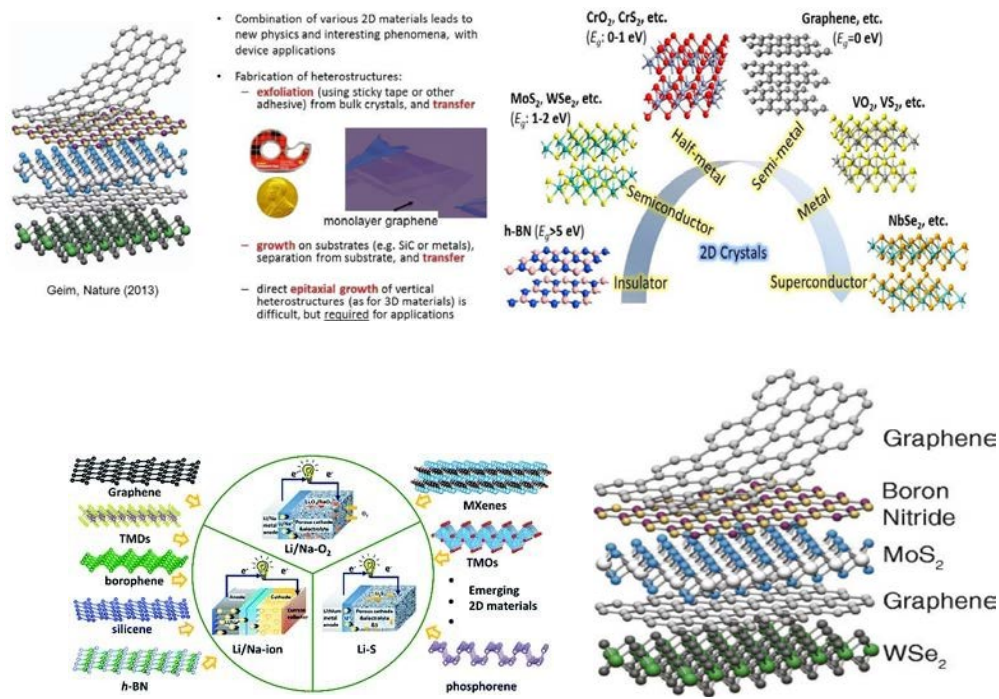

Figure 1.1: Various applications of 2D materials.

Germanene (Ge), a 2D structure of germanium, is well known for its' low buckled stable structure. This buckling provides characteristics like indirect band gap in germanane [21,22]. Just like graphene and silicene, germanene can be created by depositing germanane atoms on a substrate at high temperatures and pressure. Germanene is a zero-gap semimetal and topological insulator like graphene. Although it has high conductivity at the surface, due to its zero-band gap property, it can't be used in FET or other transistor manufacturing. The $\pi$ and $\pi^*$ bands of germanene are linear at the fermi level and are associated with massless fermions as to be the charge carriers. Though the pristine stable germanene structure involves a honeycomb shape of zero bandgaps, there is a possibility of band gap opening by an orthogonal application of electric field [23]. This is because, in the application of an external vertical field, the atoms in the buckled structures alter, and the crystal symmetry becomes inequivalent and results in a band gap formation. However, using heterostructure to evolve a bandgap from zero bandgap material has been of great interest recently. If such happens, then the fabrication of germanene that is ready to work at room temperature conditions may also be avail. Due to higher atomic numbers, the germanene is highly influenced by the spin-orbital coupling, which may be a source of the opening of the band gap at the Dirac cone, especially at the K point. The gap due to SOC may range less than 23.9 meV for germanene, referring the current spin transport via the edge can be estimated experimentally only, whereas for stanene and graphene SOC may cause a change of less than 0.05 meV and 1.55 meV, respectively [24,25,26].

Aluminum phosphide (AlP) exhibits zinc blende structure symmetry (ZB-AlP) at ambient pressure. ZB-AlP is a semiconductor material with an indirect bandgap of 2.45 eV [27]. Among the family of III–V semiconductors: the III–phosphide binary compounds crystallize in the cubic zinc-blende structure with space group (Fm3m) at ambient conditions and have an indirect band gap [28] with the exception of InP [29] By fabricating short-period superlattices (SLs) of GaP and AlP, it is possible to achieve quasi-direct band gap for certain combinations of GaP and AlP layers. These (SLs) form a new class of material, which offers the possibility of application to optical devices in the visible (green) wavelength region. This material can be used as light-emitting diodes in industrial applications. However, a 2D layer of AlP & GaP has been extracted. The most stable 2D structure of AlP is a planar hexagonal structure having a lattice constant of 3.95 Å, and for GaP, the low-buckled hexagonal structure with a lattice constant of 3.89 Å and a buckling magnitude of 0.48 Å & the band gap obtained with these structures are 2.23 eV and 1.53 eV, respectively [30]. Owing to their high band gap, they can be used as the substrate for other 2D materials having similar lattice symmetricity.

## 1.2 Literature Review

Scaling of chip size and compacting the transistor, i.e., transistor miniaturization, is going to be saturated in the upcoming years. Scientists and researchers have been searching for an alternative for the last decades. The first synthesis of a sheet of carbon which was then named graphene, showed some extraordinary features that hinted at significant mitigation for the scaling problem [1]. The problem of scaling can be reduced by using the materials used for manufacturing transistors, chips, devices. Next to graphene research was ongoing to obtain 2D materials like stanene [6,7], silicene [2,3], germanene [4,5], plumbene [8,9], 2D SiC [12],, 2D h-BN [11] etc. However, these materials, even though they possess good properties, can not instantly be used for device fabrication because of their zero band gap properties. A way was then introduced to address the situation. Hetero-bilayer of 2D materials shown a way out to introduce band gap on these zero band gap materials mentioned before. Previously, heterojunction was investigated for graphene/h-BN [31,32], stanene/h-BN [33], silicene/h-BN [35], germanene/h-BN [34], silicene/GaS [36,37], phosphorene/graphene [38], stanene/SiC [39] and other conjunctions. Some unique properties were found throughout the process that defined the applicability of these materials for optoelectronics and spintronic devices. In the last years, working with heavy zero band gap materials has become fascinating. In the process, silicene/germanene [40] was investigated where honeycomb structure of silicene and germanene were attached together, and lithography patterned lateral heterostructures were obtained that can serve in 2D electronics. In AlAs/germanene [41] hetero-structure, an indirect band gap of about 0.494 eV was obtained by Chunjian et al. Moreover, the change in band gap is obtained by their research when an external field and strain are applied to the structure. Similar findings were enlisted by the research on CdS/germanene [42] with a direct band gap of 0.644 eV, germanene/h-AlN [43] resulting in about 120 meV for four configurations and for germanene/antimonene [44] direct band gap characteristic with a moderate value of up to 391 meV were achieved. 2D germanene shows quantum spin Hall effect due to its honeycomb structure which may produce promising electronic properties. Unlike graphene, the Ge atoms on the layered germanene weak p-p interaction are usually demonstrated. Moreover, between the s and p bonds of Ge atoms, a distinct coupling exists, which may lead to a transcendence of germanene to explore new physical properties other than graphene. However, due to the indirect band gap shown by germanene hetero-structures and low order of structural stability, germanene may be restricted to narrow applications in electronics and optoelectronic devices. Research has been conducted to investigate the tunable band gap property in germanene. It was found that in the influence of spin-orbital coupling (SOC) about 23.9 meV band gap can exist at the Direct point; this in a way proves germanene to be an ideal quantum spin Hall 2D topological insulator (TI) [24]. Research conducted by María et al. [45] showed that layers of germanene could be grown on a metallic substrate like gold using dry epitaxy. Looking at the group-III phosphides (AlP, GaP) got a large indirect band gap [27], that while used as a substrate for the germanene layer, may cause distortion of germanene lattice symmetry thereby introducing a band gap. Research conducted on AlP, GaP shows that these two materials having a large indirect band are well suited for forming light-emitting diodes. GaP and AlP, when are in bulk-forming $(GaP)_m$ and $(AlP)_m$, shows that with an increase in supper lattice, their band gap decreases. Moreover, m greater than three results in a direct band gap [46]. Group III–phosphide binary compounds, when are at bulk, crystallize in the cubic zinc-blende structure with space group (Fm3m) at ambient conditions and have an indirect band gap [28] with the exception of InP [29] By fabricating short-period superlattices (SLs) of GaP and AlP, it is possible to achieve quasi direct band gap for certain combinations of GaP and AlP layers. The single-layer 2D GaP and AlP are then synthesized in the lab. The GaP being buckled its carrier are concentrated around the Dirac cones. Both AlP and GaP shows impressive Dirac cone properties and provide higher conduction capability on the surface of the bulk structure while being insulator on the interior.

Two-dimensional investigation of heterostructure using AlP and GaP are still fishing for obtaining some good and viable results. Although a few structures have been considered with these, there are no systematical illustrations of AlP and GaP influence while they are used with germanene layers by making a hetero-structure. In this dissertation, an investigation for electronic properties of 2D germanene/AlP and 2D germanene/GaP have been reported, and obtained methods and accomplished results are accreted.

### 1.3 The motivation of the Thesis

The hetero bilayer concept opened a new door for fabricating semiconductor devices with highly efficient properties. We found in literature electronic and structural properties of different heterobilayers like stanene-graphene [47], silicene- GaP [48], stanene-hexagonal boron nitride [33], germanene-hexagonal boron nitride [49] have been studied. A variation of band gap was found and opened in different layers of the hetero bilayers. The aforementioned works have provided the way to emerge band gap in zero band gap 2D layers of interest. This bandgap introduction changes the metal and semi-metal characteristics onto semiconductor characteristics, and such band, it attains a value, may result in the introduction of structure that may in future be used in devices and sensors. Research has recently focused on the germanene layer. Since germanene is a zero band gap low buckled structure, with the introduction of an external vertical field, the atoms in the buckled structures alter, and the crystal symmetry becomes inequivalent and results in a band gap formation. Again if we can use a heterobilayer taking the germanene layer on top of other 2D layers of high band gap material with a little mismatch in lattice constant, it may be avail that the influence of the wide band gap material may produce a band gap on the structure due to crystal mismatch. Since GaP and AlP, when are used at a small supercell level, are unique in their properties, they are to be used as substrate material for germanene. The phosphorous and gallium both impact quite significantly with germanium atoms. This thesis will deduce different electronic phenomena obtained by taking the different formations of Ga, P, and Ge positions in the vicinity of each other. Since, in practice, strain and interlayer distance may need to be changed, we hereby will also look for the band gap variations with such situations.

## 1.4 Objectives

The prime concern of this thesis is the theoretical investigation of structural and electronic properties of germanene and group-III nitrides (AlP, GaP) in a systematic way. However, the objectives of the research can be enlisted as follow:

- To study the structures of germanene, 2D AlP, and GaP, and to provide an investigation on the possible stacking patterns of them and to gather the basic supercell parameters.
- Observing the convergence criterion of the structures of Ge/AlP and Ge/GaP and defining the available bonding strength between them.
- Calculating the band gap and density of states for the patterns and proposing the type and magnitude of band gap expectation from Ge/AlP and Ge/GaP hetero-bilayers.
- Depicting partial density of states and charge density to obtain the atoms' orbitals that show the highest impact on the overall properties.
- To demonstrate the influence of spin-orbital coupling on the band diagram and band gap.
- To inscribe the applications of these structures in the real world and their future impacts.

## 1.5 Organization of the Thesis

The dissertation is structured in the following manner:

**CHAPTER I** encloses the general introduction of the thesis topics and materials, literature reviews of the topic, the thesis's motivation, and the thesis's objective.

**CHAPTER II** provides the basic structural and electronic properties of the materials of interest of this thesis with incorporating previous research work done on these materials.

**CHAPTER III** introduces the theoretical background and then provides a short overview of DFT framed Quantum Espresso & other associated software packages required for the thesis.

**CHAPTER IV** depicts various properties obtained from Ge/AlP and Ge/GaP hetero-structures with their comparison with previous relevant research works with brief explanations.

**CHAPTER V** wraps up the whole thesis work, enlists the outcome and unique outcome of the thesis, and paves possibilities and impact of the thesis with future research directions.

# CHAPTER II

## Structural and Electronic Properties of the Research Materials

### 2.1 Introduction

In this chapter, the structural and electronic properties of the 2D materials that are of interest to this research work are accumulated with their basic parameters. Germanene, 2D AlP & 2D GaP materials are briefly described adding the research outcomes obtained by others on them. Furthermore, the importance of using these materials is also demonstrated in a brief manner. This basis of the materials will be used for our research works outlined in the later chapters.

### 2.2 Structural and Electronic Properties of Germanene

Germanene Resembles other group-IV 2D materials, graphene and silicene, shows direct cone properties permitting the characteristics as a topological insulator [24]. Due to the buckled properties, germanene, like silicene, is influential towards an external electric field [50]. Moreover, a band gap of much larger magnitude than the graphene may get introduced in response to a break-in crystal symmetry of germanene. Again, research has shown that spin-orbit coupling also significantly influences the band gap properties of germanene [24].

#### 2.2.1  Stable Geometric Configurations of Germanene

Germanene, a single layer of germanium, was first deposited as the 2D multi-phase film upon a gold surface at high vacuum and high temperature by molecular beam epitaxy having Miller indices (111) [51]. A flat honeycomb structure was revealed using scanning tunneling microscopy (STM). Research has been conducted to determine germanene's electronic and optical properties with the Ab initio method [52]. The most stable structure of germanene is a low buckled honeycomb structure which is associated with lattice constant; a=b=4.034 Å. For free-standing germanene, DFT studies report Δ in the range of 0.64–0.74 Å [53]. Although buckling and lattice constant variant from the above may also exist, that comes with lesser stability. Free-standing bilayer germanene comes with both planar and honeycomb structures. Acun *et al.* have provided a detailed analysis of germanene lattices [55]. The structures are locally stable and optimized forms AA-stacked and AB-stacked structures, as shown in Figure 2.1. The AA-stacked structure (top panel of figure 2.1) is the planar one that possesses ~23 meV/atom resulting in more stability in nature than the AB-buckled structure (bottom panels of Figure 2.1). Thus an $sp^2$ and $sp^3$ type bonding exist in the AA and AB structure, respectively. Unlike graphene, an $sp^2$ binding in the AA structure doesn't give rise to strong intra-layer π-bonding and shows comparable strength in inter- and intra-layer interactions.

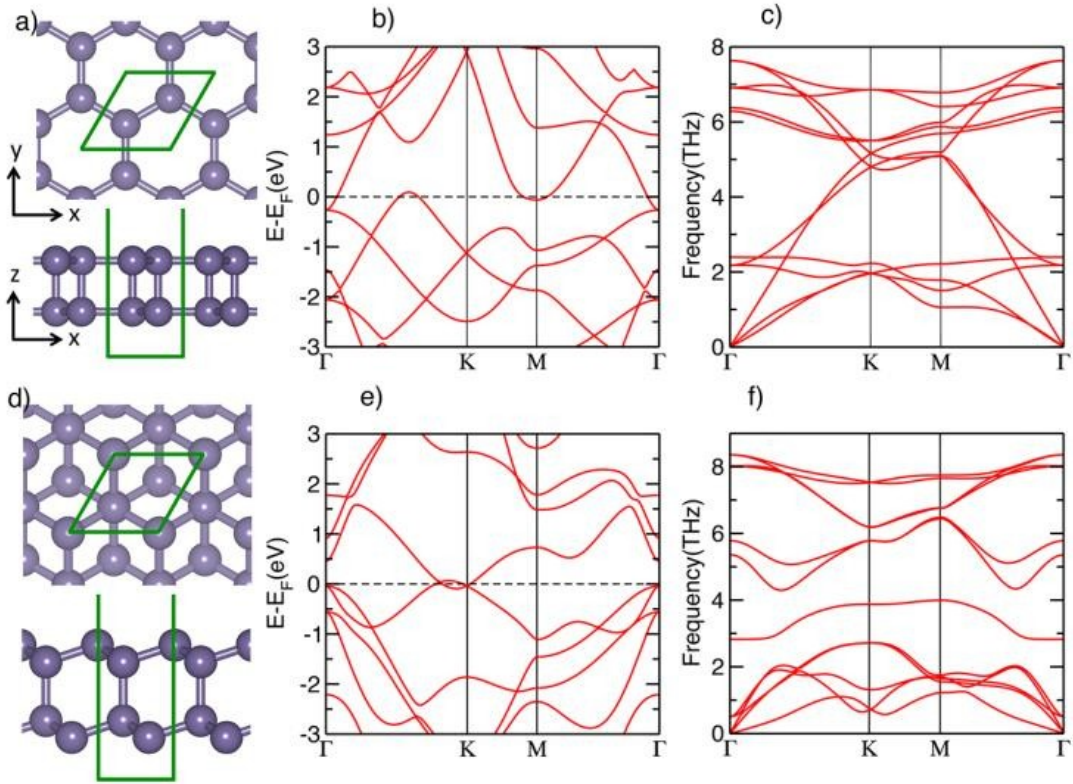

Figure 2.1: Top: (a) top and side view of the optimized AA-stacked structure of bilayer germanene (DFT calculation using the PBE/GGAfunctional); the optimized in-plane lattice constant is 4.43Å. (b) the electronic bands along with specific high symmetry directions in the 2D Brillouin zone; the zero of energy is at the Fermi level. (c) the phonon dispersions along with the same directions. Bottom: (d) top and side view of the optimized AB-stacked structure of bilayer germanene; the optimized in-plane lattice constant is 4.08Å. (e) the electronic bands and (f) the phonon dispersions of the AB-stacked structure [55].

Moreover, using LDA exchange-correlation shows that bond length for Ge-Ge for AA and AB pattern results being 2.49 Å & 2.68 Å, respectively. The local density of AA and AB structure is demonstrated by phonon spectra considering zero modes, and an imaginary frequency is depicted in figure 2.1 3(c) and (f). This absence of low-frequency optical modes like graphene shows a strong interlayer bonding [54]. Zero bands were enlisted for both the AA and AB patterns. Shyam *et al.* have shown the stable bonding length of Ge-Ge atoms with planar and buckled structures as depicted in Figure 2.2.

### 2.2.2 Electronic Properties of Germanene

Both planar and buckled germanene structure shows zero band gap around Fermi level. In corresponds to the silicene structure, the electrons in the vicinity of K and K' points on the Brillouin zone acts as massless particles. This shows the motivation of using germanene for high-speed devices. The heavy mass and atomic bond structure made germanene exhibit quantum hall effects and high magnitude change in properties after incorporating spin-orbital coupling. A band gap of small magnitude opens at the direct cone in germanene while SOC has been applied, providing its application as topological insulators. It is reported that about 23.9 meV of the band gap is opened by this process [24]. However, the properties of the germanene can be modified by taking the interaction of the substrate with germanene [57,58]. Research shows that applied strain may cause a change in buckling length, and since the sub-lattice symmetry is not preserved like graphene, a band gap emerges [59]. With applied strain and breaking of germanene symmetry band gap position in a hetero-structure shows a variation from indirect to direct and shift of band gap from one significant first Brillouin Dirac point to another appears with change in band gap levels.

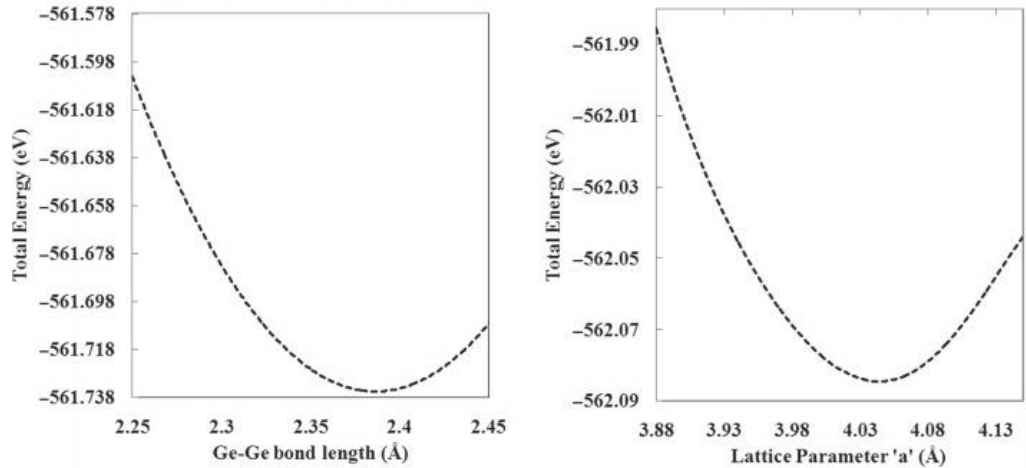

Figure 2.2: Left: Energy variation with the bond length for planar germanene for fixed lattice a = 4.130 Å, Right: Energy variation with the bond length for planar germanene for fixed lattice a = 4.130 Å [56].

## 2.3 Structural and Electronic Properties of 2D-AlP and 2D-GaP

Group III–phosphide binary compounds, when are at bulk, crystallize in the cubic zinc-blende structure with space group (Fm3m) at ambient conditions and have an indirect band gap [28] with the exception of InP [29] By fabricating short-period superlattices (SLs) of GaP and AlP, it is possible to achieve quasi direct band gap for certain combinations of GaP and AlP layers. The single-layer 2D GaP and AlP are then synthesized in the lab. The GaP being buckled its carrier are concentrated around the Dirac cones. Both AlP and GaP shows impressive Dirac cone properties and provide higher conduction capability on the surface of the bulk structure while being insulator on the interior. Research conducted focusing on band alignment of 2D semiconductors for designing hetero-structure with moment space matching provides a detailed view of AlP and GaP and their lattice parameters [30]. They investigated materials using both the GGA-PBE and HSE methods and accredited the band gap outcomes and position of the bandgap at the 1st Brillouin zone. The lattice constant for stable planar AlP was found as a=b=3.95 Å and bond length d=2.28 Å. In the case of GaP stable structure for conducting hetero-structure with a=b=3.89, bond length

d=2.30 Å and buckling length of about 0.48 Å. Figure 2.3 and 2.4 shows the band diagrams obtained for AlP and GaP using the HSE method, respectively.

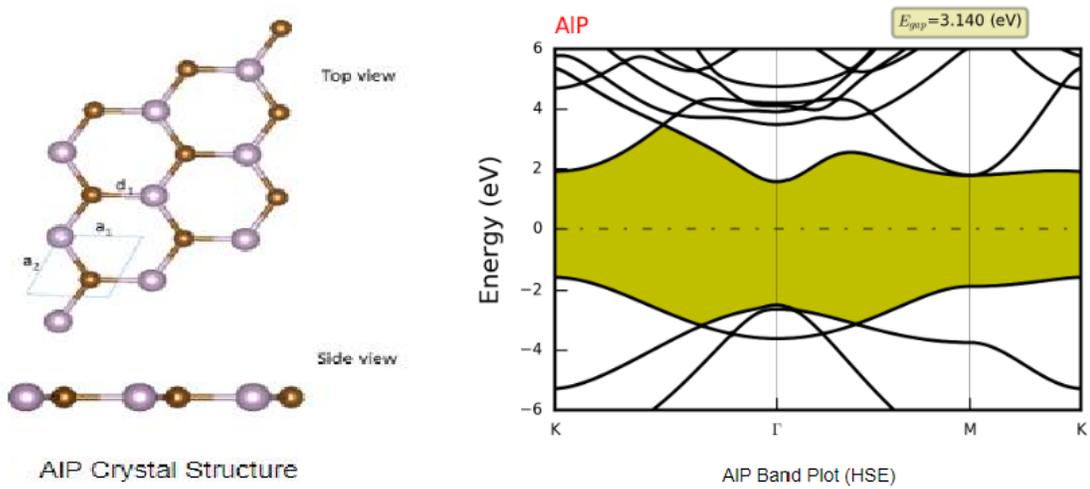

| | PBE | HSE |
|---|---|---|
| Valence Band Maximum | -5.32 eV | -5.76 eV |
| Conduction Band Minimum | -3.09 eV | -2.62 eV |
| Band Gap | 2.23 eV | 3.14 eV |

Figure 2.3: 2D AlP structure with band diagram [30].

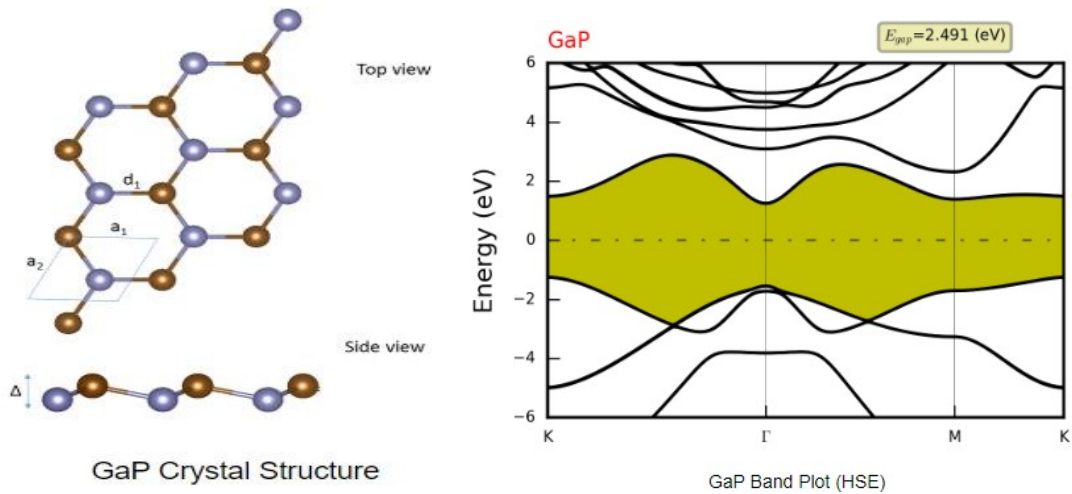

| | PBE | HSE |
|---|---|---|
| Valence Band Maximum | -5.46 eV | -4.29 eV |
| Conduction Band Minimum | -3.93 eV | -1.80 eV |
| Band Gap | 1.53 eV | 2.49 eV |

Figure 2.4: 2D GaP structure with band diagram [30].

It is observed that using the HSE method provides a higher band gap prediction than that obtained using the PBE technique. An indirect band gap is shown between gamma and K points in both cases. Owing to these high band gaps in this thesis, they are going to be used as the substrate for germanene growth. Due to the closer lattice symmetry between the germanene/2d-GaP and germanene/2d-AlP. But in practice, supercells are utilized to depict the actual observations. In the case of GaP-AlP (001), it was found that transition from the valence-band maximum to the conduction-band minimum state at the $\Gamma$ point is much stronger in an even number of super lattice periods. By tuning the strain and lattice symmetricity, their band gap can be made versatile between indirect to direct in between the first Brillouin zone point, as shown in figure 2.4.

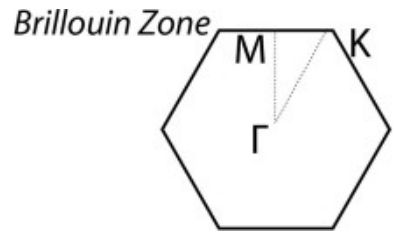

Figure 2.5: First Brillouin zone for hexagonal structures [30].

# CHAPTER III

## Computational Details

### 3.1 Introduction

This chapter of the research manuscripts amalgamates an extensively detailed description of the theoretical background and simulation schemes utilized for the computation. In addition to the above-mentioned, Quantum Espresso [60] simulation package and corresponding *Ab initio* codes for the PWSCF suite are also added; these codes are the base to produce the outcomes that have been analyzed in the subsequent chapters.

### 3.2 Density Functional Theory (DFT) Framework

DFT is, in general, defined as a quantum mechanical modeling method used for the investigation of many-body systems. The density functional theory title is formulated due to the fact that it accounts for a space-dependent electron density. It is termed as to be the most popular method in avail to address computational physics and chemistry. DFT function scheme directly follows the equation of Schrödinger and is termed being First principle or Ab Initio method. For a quantum, mechanical system, ground state energy is being utilized while considering the development of an ab initio scheme. The ground state represents the most stable state of a system, which generally has the lowest energy. Schrödinger equation to find the ground state for a collection of atoms can be defined by

$$H\psi(\{r_i\}, \{R_I\}) = E\psi(\{r_i\}, \{R_I\})$$

This Hamiltonian shows the basic relationships between the wave function and energy states. Starting from the tight-binding model of solid to the quantum mechanical calculations, this forms the basic level of understanding. Furthermore, effective mass —is usually calculated from its basis.

Here,

- $H$ stands for the Hamiltonian of a quantum mechanical system, $\mathbf{H} = \mathbf{T} + \mathbf{V}_{\text{columb}}$. Where $T$ refers to the kinetic energy operator,

$$\mathbf{V}_{\text{columb}} = \frac{Q_i Q_j}{|r_j - r_i|}$$

  Where, $Q_i$ and $Q_j$ are pair of charged particles.

- $\psi$ stands for the wave function of an atom.

- $r_i$, $R_I$ represents the position of the electron and nucleus of an atom.

- $E$ represents the energy eigenvalue.

The approximation investigates the many-body Schrödinger equation for an atom containing the nuclei and electrons by the Born Oppenheimer, where the atomic wave function consideration is decoupled into nuclei and electron parts contribution.

### 3.2.1 Born Oppenheimer (BO) Approximation

BO approximation is a basic quantum mechanical approach that describes the quantum states of molecules. BO approximation paves the way to separate the motion of electron from the motion of nucleus in a molecule, taking the consideration that electronic wave function only depends upon the position of nucleus rather than their velocities, and from the speedy electron, the nucleus motion sees smearing out potential. The consideration enclosed utilized the fact that the mass of the nucleus is way much heavier than the electron; thus, electrons are way too fast-moving particles than nuclei, & dynamics of atomic nuclei and electrons can be separated. Due to this, the motion of nuclei is usually neglected, and consideration of electronic waves is only taken based on the assumed static position of the nuclei, neglecting the motion associated with the nuclei. By dint of BO approximation, the wave function can be decoupled as following

$$\psi_{\text{atomic}}\left(\{r_i\}, \{R_I\}\right) = \psi_{\text{nuclei}}\left(\{r_{ij}\}\right) * \psi_{\text{electron}}\left(\{R_I\}\right)$$

Here, $m_{nuclei} \gg m_{electron}$ is the consideration to neglect the motion of nuclei. This decoupled arrangement of the nuclei and electron in the way to explain the many-body Schrödinger equation is further aggrandized by converting the N-particle equation into N-single particle equations by means of Hartree-Fock (HF) approximation.

### 3.2.2 Hartree-Fock (HF) Approximation

The separation of an electron from the nuclei results in a system of N-electron many-body system and can be described by Schrödinger equation

$$\boldsymbol{H}\psi_i(1, 2, \ldots N) = \boldsymbol{E}\psi_i(1, 2, \ldots N)$$

Representing the terms associated with the N-electrons in the i[th] atom. The degrees of freedom for particles in the braces (1, 2, …., N) depend on the position and spin of the electron in general. The electronic Hamiltonian then can be divided into constituent terms as

$$\boldsymbol{H} = -\frac{\hbar^2}{2m}\sum_{i=1}^{N}\nabla_i^2 + \sum_{i=1}^{N}V_{ext}(r_i) + \sum_{i=1}^{N}\sum_{j>1}^{N}U(r_i, r_j)$$

Considering $r_i$ and $r_j$ being the position of $i^{th}$ and $j^{th}$ electron within the N electrons of the system. The $1^{st}$ term on the right-hand side represents the kinetic energy of the $i^{th}$ electron in the system; the $2^{nd}$ part gives the potential energy between the Coulomb attraction between the $i^{th}$ electron and protons of the nuclei under the system, and the final part shows the influence of $j^{th}$ electron by other electrons altogether. Considering heavier atoms, the number of electron particles increases, and if their degree of freedom is higher, the system becomes complex to investigate. HF approximation shows an approach where each electron feels the average coulomb repulsion by the other electrons of the system, which makes possible the separation of N-electron system to N-one electron system as following

$$\psi_i(r_1, r_2, \dots \dots r_N) = \psi_i(r_1) * \psi_i(r_2) * \dots \dots * \psi_i(r_N)$$

Though the HF approximation simplifies the many-body system to many one-body systems, it often ends up in ambiguities. The reason behinds the underlying fact is that the HF approximation does not consider the probability of the correlation energy contribution by a third electron in the vicinity of two electrons exchange-correlation. Moreover, due to the fact that while calculating the exchange contribution Pauli Exclusion Principle is utilized, the contribution by similar fermion with similar quantum states is also excluded. Thus the evolved exchange-correlation functional by the HF approximation gives bad results while calculating system energy utilizing this. The density functional theory (DFT), many-body perturbation theory (MBPT), configuration interaction (CI), etcetera provides a way to calculate the correlation energy, which then is explicitly added to the system energy obtained by the HF approximation. The electron density for the N-electron system is given by

$$n(\boldsymbol{r}) = \psi^*(r_1, r_2 \dots \dots r_N) * \psi(r_1, r_2 \dots \dots r_N)$$

The electron density in terms of the individual electron wave functions can be obtained as

$$n(\boldsymbol{r}) = 2 \sum_{i=1}^{N} \psi_i^*(\boldsymbol{r}) \psi_i(\boldsymbol{r}) = \sum_{i=1}^{N} |\psi_i(\boldsymbol{r})|^2$$

Hohenberg and Kohm (HK) considered the problem and provided a way to increase the accuracy and provided a way to formulate the many-electron wave function to electron density.

### 3.2.3 Hohenberg and Kohn (HK) Theorems

HK theorem is termed being the successful first step in consideration of DFT and the heart of DFT. In the absence of relative velocities, the Hamiltonian for a many-electron system can be given by the relation

$$H = -\frac{\hbar^2}{2m} \sum_{i=1}^{N} \nabla_i^2 + \sum_{i>j}^{N} \frac{1}{|r_i - r_j|} + \sum_{i=1}^{N} v(r_i)$$

Where the $1^{st}$, $2^{nd}$ & $3^{rd}$ terms represent the $i^{th}$ electron's kinetic energy, the complexity of the problem caused by the degree of interaction between the atoms, and external potential, which involves all effects external to electrons, including Coulomb potential between moving electron and static Nuclei and from other possible fields, respectively. HK is an exact DFT theory for the many-body system; it refers to the system of many electrons that are influenced under an external field $V_{ext}(\mathbf{r})$, $\mathbf{r}$ being the separation of electron and nucleus. HK demonstrates two theorems as:

**First Theorem:**

The external energy and the whole energy is unique functional of the ground state electron density. This, in terms, shows the appliance of ground-state electron density for obtaining all the system properties. Energy functional of ground-state electron density E[n(r)] can be expressed in terms of external potential by

$$E[n(\mathbf{r})] = \int n(\mathbf{r}) V_{ext}(\mathbf{r}) dr + F[n(\mathbf{r})]$$

Here, F[n(**r**)] is an unknown universal HK functional of electron density which can also be applicable to zero external energy situations; on the right-hand side, the $2^{nd}$ term shows the system-dependent energy, and the $1^{st}$ part gives the energy associated with the universal functional. This theorem also gives that the ground state energy $E_0$ is a unique function of the electron density $n(\mathbf{r}_0)$. Again

$$F[n(\mathbf{r})] = T[n(\mathbf{r})] + U[n(\mathbf{r})]$$

Where, T[n(**r**)] and U[n(**r**)] represents the total kinetic energy functional and potential energy functional extracted from electron-electron interactions, respectively.

**Second Theorem:**

The energy obtained by the first theorem is the function of electron density. The electron density for which minimum energy of the overall function is obtained is the true ground-state electron density defined by the relation

$$E[n(\mathbf{r})] > E_0[n_0(\mathbf{r}_0)]$$

$$E_0[n_0(\mathbf{r}_0)] = minE[n(\mathbf{r})]$$

Thus, exact ground energy can be calculated by utilizing the total energy of the system and the exact ground-state electron density. Any variation on the ground state density changes the energy obtained, which by definition is greater than the exact ground state value. HK theorem in this way can theoretically be used for finding the ground state energy, which may then be used for the computation of other system properties. However, the limitation of the HK method shows that in practice, there is no defined route to the formulation of exact ground electron density of state.

Although addressed by the HK methods thus is actually impractical in the way that they were unable to give an exact solution to the problem. However, an easier scheme proposed by Kohn and Sham (KS) scheme is utilized in practice [61].

### 3.2.4 Kohn-Sham (KS) Formulation

The HK theorem considered and looked for ground-state electron density for each particle individually. However, KS formulated a method that considered the ground state densities of all the particles as a single set and utilized it to find out the whole system's exact ground state energy density. According to the KS formulation, the ground state electron density can be located by considering the ground-state electron density of a fictitious system consisting of non-interacting particles. These non-interacting individual particles can be addressed by a set of individual particle Schrödinger equations that can be solved numerically to find the ground-state electron density and exact ground state energy. KS changed the total energy onto two terms, primarily

$$E[n(\boldsymbol{r})] = E_{known}[n(\boldsymbol{r})] + E_{xc}[n(\boldsymbol{r})]$$

$$E_{known}[n(\boldsymbol{r})] = \int n(r)V_{ext}(r)dr + \frac{1}{2}\iint \frac{n(\boldsymbol{r})n(\boldsymbol{r}')}{|\boldsymbol{r}-\boldsymbol{r}'|}d\boldsymbol{r}d\boldsymbol{r}' + T[n(\boldsymbol{r})]$$

Here, for the right-hand side of the second known portion of the total energy

- The first part refers to the energy associated with the system in the external potential.

- The second part gives the classical Coulomb interaction energy due to electron energy density ($|\psi|^2$ or $\psi\psi*$) and is termed Hartree energy $V_H(\mathbf{r})$.

- The third term represents the kinetic energy associated with the individual particles.

$E_{xc}[n(\mathbf{r})]$ is the exchange-correlation energy functional that combines all the many-body electron-electron interactions in the system and, in general, is the difference between exact total energy and other known energy values. By definition, the exchange-correlation energy must contain the self-interaction term that is equal to the Hartree energy and both equal out. The form of exchange-correlation energy is unknown, and approximations are utilized while calculating DFT.

The ground-state electron density is calculated by assuming a set of individual no-interacting particles equation, according to KS is given by considering individual particle KS orbital $\psi_i(\mathbf{r})$ and effective KS Hamiltonian $H_{ks}^{eff}$ as

$$H_{ks}^{eff}\psi_i(\boldsymbol{r}) = \varepsilon_i\psi_i(\boldsymbol{r})$$

The relation between KS orbitals and electron density shows the relation

$$n(\boldsymbol{r}) = \sum_{i=1}^{N} |\psi_i(\boldsymbol{r})|^2$$

If the ground state density is exact, the KS effective Hamiltonian can be given by

$$H_{ks} = -\frac{1}{2}\nabla^2 + v_{ks}(\boldsymbol{r})$$

Where $V_{ks}(\boldsymbol{r})$ is the KS potential and is defined as

$$v_{ks}(\boldsymbol{r}) = v_{ext} + v_H(\boldsymbol{r}) + v_{xc}(\boldsymbol{r})$$

The above four equations are used iteratively, starting with an initial ground density of state $n_0(\boldsymbol{r})$ that is used through equations 2nd, 3rd, 4th, and 5th successively to get KS potential $V_{ks}(\boldsymbol{r})$ This $V_{ks}(\boldsymbol{r})$ is then utilized through equations 1st and 2nd successively, resulting in a new density of state $n_0(\boldsymbol{r})$. This value is compared with the initial ground density of state, and if the value matches the initial value, the exact ground density of states is obtained; if a mismatch occurs, a new trial density of state with minimization of total energy value is utilized thus moving towards the lower energy level indicating the ground state. This method is called Self-consistency in DFT calculation, and a convergence threshold is prescribed before running the computation. The exchange correlation is usually approximated in two ways; Local Density Approximation (LDA) and Generalized Gradient Approximation (GGA).

### 3.2.5 Local Density Approximation (LDA)

LDA refers to the situation where the electron densities are considered locally such that exchange-correlation energy for each point of the system is the same as that of a uniform electron gas of the same density throughout the system. This approximation holds true for a system with slowly varying densities. The homogenous electron gas is referred to as jellium. The exchange-correlation energy $E_{XC}[n(r)]$ can then be calculated using LDA by

$$E_{XC}^{LDA}[n(\boldsymbol{r})] = \int n(\boldsymbol{r})\varepsilon_{XC}(n)d^3\boldsymbol{r}$$

Although the LDA is a simple method for obtaining exchange-correlation energy, it is significantly characterized by different drawbacks. Determination of eigenvalues with LDA is not always hold accurate in addition to the problems

    a. Derivative terms may be discontinuous.

    b. Provides good results with ground state situations only, and while excited states are considered, LDA outputs poor results.

    c. LDA exhibits self-interaction difficulty.

    d. LDA provides a bad prediction of metal-oxides.

    e. Spin and angular momentum are not sufficiently large in LDA.

f.  LDA underestimates semi-conductor as well as insulator bandgap.

g.  Results shown by LDA for bulk moduli and cohesive energies are insignificantly very high.

However, the LDA paves the way for the construction of more sophisticated approximations like the GGA and hybrid functionals since it is associated with Homogeneous Electron Gas (HEG) for non-varying density that may result in an exact exchange-correlation (EXC) energy.

### 3.2.6    Generalized Gradient Approximation (GGA)

The aforementioned LDA is too simple and erroneous for real systems; thus, Hohenberg and Kohn prescribed an extension to the LDA known as the Gradient Expansion Approximation (GEA)*. The GEA considers a series expansion of increasingly higher-order density gradient terms, which was then implemented to calculate atoms and molecules, and the result was bad and frustrating. Despite the disappointing result found for first-order GEA, it forms the basis for the most utilized and popular GGA exchange-correlation functional in condensed physics. In a situation where electron density changes rapidly, the LDA fails, but in GGA, in addition to the local electron density, the magnitude of the gradient of electron density is considered. Thus, the overall performance increases. An analytical function known as enhancement factor $F_{XC}[n(r), \nabla_n(r)]$ is usually used to modify the LDA to obtain GGA

$$E_{XC}^{GGA}[n(r)] = \int n(r) \varepsilon_{XC}^{homo} F_{XC}[n(r), \nabla_n(r)] d^3r$$

The use of GGA results in good outcomes for molecular geometries, and more accurate ground-state energies have been achieved. In present days meta-GGA, an upgraded edition of GGA, provides more efficacy than the GGA. These form the basis of DFT calculation for all condensed materials.

### 3.3    Pseudo-Potential

Coulomb potential between the nucleus and electron can be given by:

$$V_{nuc} = \frac{-Ze}{r}$$

Here, *Ze* is the nuclear charge, and *r* stands for the distance between nucleus and electron. Because of the tightly bound core orbitals with the nucleus, and the highly oscillatory nature of the valence electrons, demand that a high value of $E_{cut}$ is required. This creates a computational problem to reach the Kohn-Sham equation's solution. Since most of the physical properties of solid depend solely on the valence electron and the core electrons behave almost independently in the environment, it is possible to partition the electrons between core and valence state. The core electrons and ionic potential can be replaced with a pseudopotential. Kohn-Sham wave functions can be expanded to a summation of plane waves $\phi_{i,G}$ using this pseudopotential. As a result, an orthogonal plane-wave basis is created by leading to the following equation:

$$\psi_i(\boldsymbol{r}) = \sum_G C_{i,G}\,\phi_{i,G}$$

In Frozen Core Approximation (FCA), the interaction among the valence electrons is taken into consideration in DFT that faces an effective pseudo-potential, which is more flexible than the

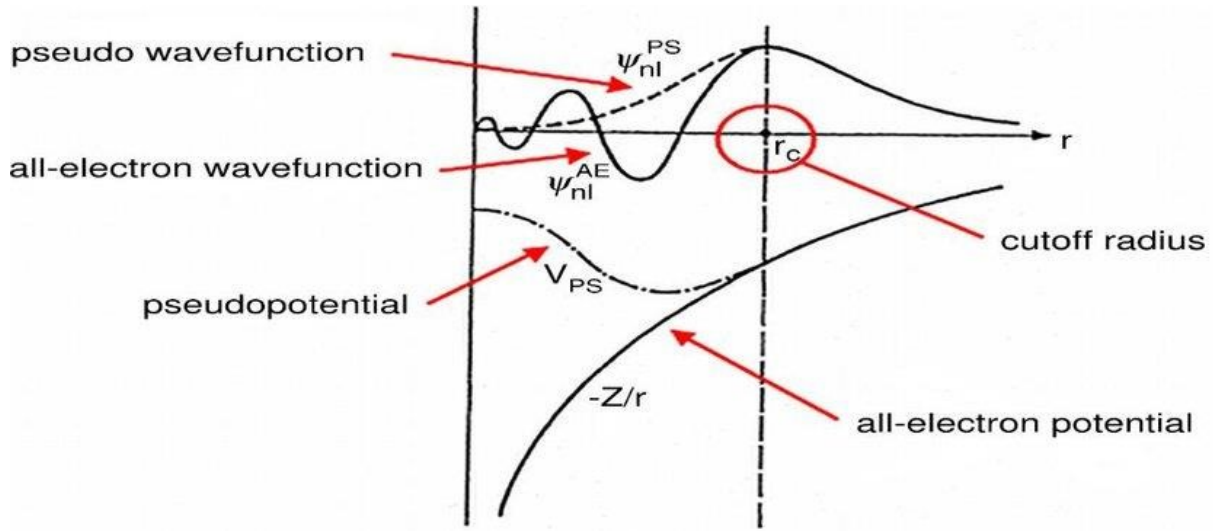

Figure 3.1: Depiction of the basic function of pseudopotential

Coulomb potential of the nucleus. In this way, pseudopotential is constructed such that the pseudo wave function exhibit no radial nodes within the core region. This is depicted in Figure 3.1.

In Figure 3.1:

- $\psi_v = \psi_{ac}$ represents the all-electron-wave function without FCA
- $\psi_v = \psi_{ac}$ represents pseudo-potential wave function in FCA
- $V_{pseudo}$ represents the pseudo-potential that generate the pseudo-potential wave function
- $r_c$ represents the cut-off radius under which FCA is utilized.

### 3.3.1 Properties of Pseudo-Potential Wave Functions ($\psi_{pseudo}$)

Pseudo-potential possesses some significant properties based on their application, and many types of pseudo-potential can be used in different situations. For example, transferability of pseudo-potential is a sublime property which means that the same pseudo-potentials can be used in various chemical situations per element. However, for the pseudo-potential to be transferable, it is necessary for the pseudo-potential to have similar scattering features as the all-electron potentials.

According to quantum mechanics, this feature is connected with the logarithmic derivatives, which are written as:

$$\frac{d}{dr}\ln\left(\psi_{ae}(\boldsymbol{r})\right) = \frac{d}{dr}\ln\left(\psi_{\text{pseudo}}(\boldsymbol{r})\right)$$

Identical pseudo-potential will be obtained such that beyond the cut-off radius, the pseudo wave functions are as same as the all-electron-wave defined by

$$\psi_{ae}(\boldsymbol{r}) = \psi_{\text{pseudo}}(\boldsymbol{r})$$

For an ideal pseudo-potential, its eigenvalues are independent of the quantum numbers ($l$). Henceforth,

$$\varepsilon_l^{ae} = \varepsilon_l^{p\,\text{seudo}}$$

Again, non-conserving properties are shown by an ideal pseudo-potential, which indicates that inside the sphere of the cut-off radius ($\boldsymbol{r}_c$), the total charge associated with each wave function might resemble the charge of the all-electron wave function. This can be written as:

$$4\pi\int_0^{r_c}|\psi_{ae}(\boldsymbol{r})|^2\boldsymbol{r}^2 d\boldsymbol{r} = 4\pi\int_0^{r_c}\left|\psi_{\text{pseudo}}(\boldsymbol{r})\right|^2\boldsymbol{r}^2 d\boldsymbol{r}$$

### 3.3.1   Soft and Hard Pseudo-Potentials

With a view to investigating whether a pseudo-potential is a soft or a hard one, the following figure is usually considered where $r_c$ is the cut-off radius.

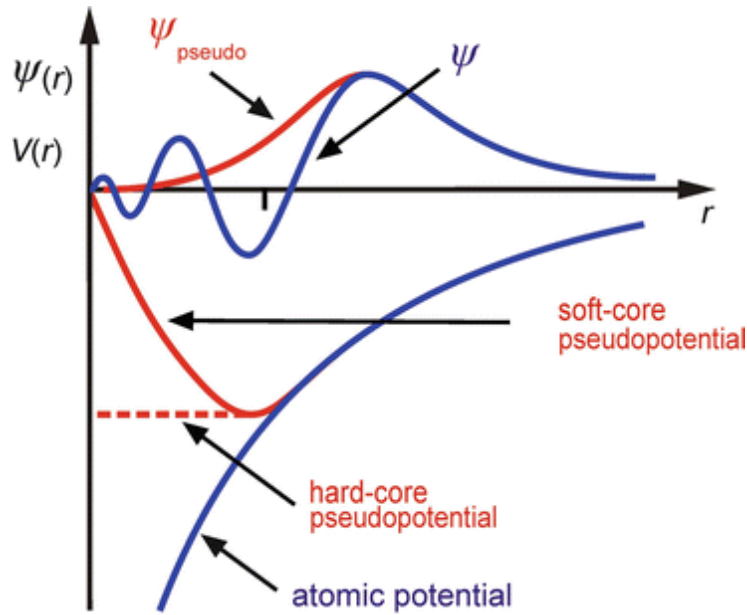

The value of cut-off radius ($r_c$) can be obtained by taking the following considerations :

- For the outer-most node of all-electron-wave-function $r_c > r$.

- $r_c$ lies within the outermost node of the all-electron wave function, and in the case of the outermost node, it is the highest wave function value.

In the case of $r_c$ being much smaller, a hard pseudo-potential is applied. Hard pseudo-potential generally to achieve convergence extend the Kohn-Sham wave functions over a huge sum of plane waves. Again, a larger $r_c$ value results in poor transferability. Therefore, care needs to be taken while considering the value of $r_c$ and is to be chosen in a way that the pseudo-potential would remain practically 'soft' (large value of cut-off radius). And to exacerbate the operation with the soft pseudo-potential, the cut-off energy of the plane wave basis ($E_{cutwfc}$) isn't taken too high while considering DFT calculations. Anyway, DFT simulation suites generally pre-calculate three terms of $V_{eff}(r)$, which is followed by their addition, and then stocks the result in pseudo-potential files for user access.

Figure 3.2: The cut-off radius position which decides the type of pseudo-potential

### 3.3.2 Pseudo-Potentials Naming Convention

It is necessary to have good knowledge about the pseudo-potentials to perform LDA or GGA computation in QE. Selecting a perfect pseudo-potential to apply it in the input for LDA or GGA is essential. Using a different class of pseudo-potentials for different materials within the same compound may produce the wrong data. In this section of the dissertation, a little about the pseudo-potential naming convention will be discussed, which will help one to choose the correct pseudo-potential.

Pseudo-potentials (PP) in the Quantum Espresso are named using the Unified Pseudo-potential Format, UPF. For the UPF PP files to name, the following criterion is fulfilled:

**material.explanation.UPF**

**'material'** is the name of the corresponding element. For example, if the material is Carbon, then this part will be replaced by C.

The 'explanation' consists of several fields, which are separated using a dash (_). Four fields are described below:

- **Field # 1:** This field is used if the pseudo-potential is relativistic. If not, this field is left empty.
  rel: full-relativistic

- **Field # 2:** Second field is a mandatory field, which corresponds to the exchange-correlation functional suited for the pseudo-potential.
  o **pbe:** Perdew-Burke-Ernzerhof, GGA exchange correlation.

- o **pz:** Perdew-Zunger, LDA exchange correlation.
- o **vwn:** Vosko-Wilk-Nusair, LDA exchange-correlation.
- o **coulomb:** Coulomb bare –Ze/r potential.
- o **blyp:** Becke-Lee-Yang-Parr, GGA exchange-correlation.
- o **pw91:** Perdew-Wang 91 gradient corrected functional, GGA exchange-correlation.

- **Field # 3:** The third field is not a mandatory field. It may cover one or more than one of the following:
  - o **s:** valence s state
  - o **p:** valence p state
  - o **d:** valence d state
  - o **f:** valence f state
  - o **n:** nonlinear core-correction

- **Field # 4:** Being the last field, it conveys information about the author or the version of the pseudo-potential or both. If information about both the version and the author is given, then this field comprises the PP version first, which is followed by the author's info.
  - o **mt:** Martins-Troullier
  - o **ae:** All-Electron (no pasteurization)
  - o **vbc:** Von Barth-Car (direct fit)
  - o **bhs:** Bachelet-Hamann-Schlueter and derived
  - o **rrkj:** Rappe Rabe Kaxiras Joannopoulos (norm-conserving)
  - o **van:** Vanderbilt ultra-soft
  - o **bpaw:** Projector Augmented Wave (original Bloechl recipe)
  - o **kjpaw:** Projector Augmented Wave (Kresse-Joubert paper)
  - o **rrkjus:** Rappe Rabe Kaxiras Joannopoulos (ultra-soft)

### 3.4  Plane-Wave Cutoffs and K-Point Sampling

While performing DFT pseudo-potential computation, usually two critical value needs to be set up. The energy cut-off ($e_{cutoff}$) value provides the plane wave expansion, and a k-mesh grid is utilized that is used to sample the continuous integral on different symmetrical points on the 1st Brillouin zone. The higher the value of cut-off and k-mesh, the accuracy of the computation increases. But an increase in the cut-offs takes much more computational force, thus taking more computational time. Usually, a balance is taken so that less computational time with higher accuracy is obtained.

### 3.4.1 Convergence and Plane-Wave Basis

DFT plane-wave calculation includes a lot of systems through the use of periodic boundary conditions. Therefore, it is necessary to go with the Bloch theorem with a view to dealing with these systems. The Bloch theorem is:

$$\psi_{nk}(r) = \exp{(ik.r)}\mu_{nk}(r)$$

In the above equation $\psi_{nk}$ represents the electronic wave function which is written in the following:

$$\mu_{nk}(r) = \sum_{G} C_G \exp{(iG.r)}$$

Here, $G$ makes a summation of all lattice vectors, although actually, it might be shortened at a certain point. $C_G$ represents the expansion coefficients. In the QE simulation package, the plane waves are the basic functions. Plane waves well-suited for direct lattice boundary conditions and having periodicity are chosen by restraining the plane-wave expansion to the discrete set of $G$ vectors. $G$ vectors are the multiples of primitive lattice vectors, and they will not be a fraction. Therefore, it is necessary to take the plane-wave basis to the boundary of an infinite number of $G$ vectors. Practically, instead of this, the plane waves are truncated at a certain point which is the plane-wave cut-off. The cut-off energy in units of Rydberg (Ry) or electron-volt (eV) corresponds to the highest value of kinetic energy $(k + G)$. Again, the electronic states of the Brillouin zone are dependent on the Brillouin zone's k-points. If the number of k-points increases, the number of repeated unit cells (N) will also increase.

### 3.4.2 K-Point Sampling and Convergence

In the Bloch theorem, it is necessary to solve the Kohn-Sham equation at every point of the Brillouin zone. This requires the diagonalization of self-consistent interactive $N{\times}N$ matrix; here, $N$ stands for the number of basis functions. Integration has to be carried over the first Brillouin zone. In order to perform practical computation, the integration carried along the first Brillouin zone is exchanged with a fixed sum over discrete $k$ points. The $k$-points are taken in order to repeat a very good approximation result on integration. One must be careful to adopt sufficient k-points to meet the convergence to the system's overall energy. To allow sampling of the Brillouin Zone on the basis of a number set of $k$ values, a 'special-points' scheme may be utilized. This type of scheme is very useful to sample the Brillouin zone as they offer an effective way to perform integration on the periodic functions.

### 3.4.3 Monkhorst and Pack Grid for Brillouin Zone Sampling

DFT plane-wave calculation includes a lot of systems through the use of periodic boundary conditions. Therefore, it is necessary to go with the Bloch theorem with a view to dealing with these systems. The Bloch theorem is:

$$\psi_{nk}(r) = \exp{(ik.r)}\mu_{nk}(r)$$

In the above equation $\psi_{nk}$ represents the electronic wave function which is written in the following:

$$\mu_{nk}(r) = \sum_G C_G \exp{(iG.r)}$$

Here, $G$ makes a summation of all lattice vectors, although actually, it might be shortened at a certain point. $C_G$ represents the expansion coefficients. In the QE simulation package, the plane waves are the basic functions. Plane waves well-suited for direct lattice boundary conditions and having periodicity are chosen by restraining the plane-wave expansion to the discrete set of $G$ vectors. $G$ vectors are the multiples of primitive lattice vectors, and they will not be a fraction. Therefore, it is necessary to take the plane-wave basis to the boundary of an infinite number of $G$ vectors. Practically, instead of this, the plane waves are truncated at a certain point which is the plane-wave cut-off. The cut-off energy in units of Rydberg (Ry) or electron-volt (eV) corresponds to the highest value of kinetic energy ($k + G$). Again, the electronic states of the Brillouin zone are dependent on the Brillouin zone's k-points. If the number of k-points increases, the number of repeated unit cells (N) will also increase.

Monkhorst-Pack grid [62] is an efficient and balanced scheme to select a K-point mesh to sample the Brillouin Zone. In this scheme, there will be a homogeneous distribution of the K-points in the Brillouin Zone. The columns or the rows will be parallel to the reciprocal lattice vectors spanning the Brillouin zone. The Brillouin Zone sampling of a hexagonal lattice with the Monkhorst-Pack grid is presented in Figure 3.4. The reciprocal lattice vectors (b1, b2, and b3).

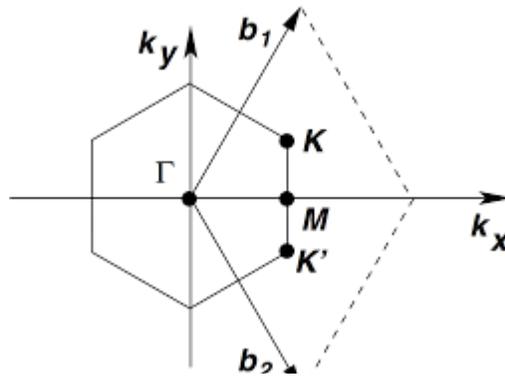

Figure 3.3: Monkhorst-Pack grid scheme to sample the Brillouin Zone

Which is connected with the origin of the coordinate system that spans the Brillouin Zone. In origin, the Brillouin Zone will rest, as stated by this scheme. The entire Brillouin zone will be tiled by smaller polyhedra of the identical shape of the Brillouin zone. A rectangular point grid with dimensions $M_x \times M_y \times M_z$ will space regularly all over the Brillouin zone.

## 3.5 Quantum Espresso and DFT

### 3.5.1 Self-Consistent Field Cycle

The Kohn-Sham equations offer a practical way to determine the ground state energy and particle density of a system. These computations mainly adopt the Schrödinger equation to accomplish self-consistent calculation.

The procedure to solve the Kohn-Sham equations for the electronic charge density is illustrated in Figure 3.4. The procedure starts by making a guess on the electronic charge density. The effective

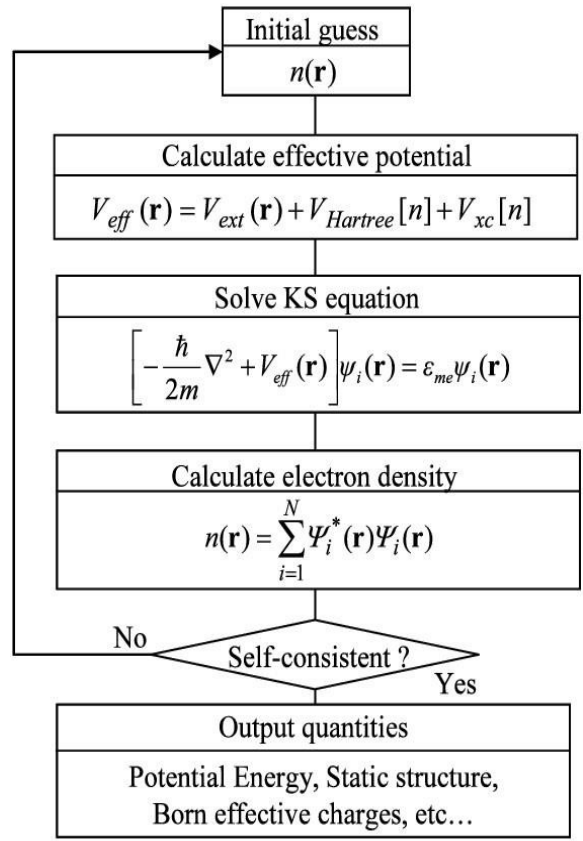

Initial guess

$n(\mathbf{r})$

Calculate effective potential

$V_{eff}(\mathbf{r}) = V_{ext}(\mathbf{r}) + V_{Hartree}[n] + V_{xc}[n]$

Solve KS equation

$\left[ -\frac{\hbar}{2m}\nabla^2 + V_{eff}(\mathbf{r}) \right]\psi_i(\mathbf{r}) = \varepsilon_{me}\psi_i(\mathbf{r})$

Calculate electron density

$n(\mathbf{r}) = \sum_{i=1}^{N}\Psi_i^*(\mathbf{r})\Psi_i(\mathbf{r})$

No — Self-consistent ? — Yes

Output quantities

Potential Energy, Static structure, Born effective charges, etc…

Figure 3.4: Self-Consistent Field Flowchart

potential ($V_{eff}$) is then determined using the charge density. The determination of the effective potential shows the way to find the solution of the system's ground state in the Kohn-Sham equations. The self-consistent field (SCF) cycle utilizes the Kohn-Sham equations' solution to have a new electron density. After calculating the new electron charge density, the system will check if the convergence criterion has been achieved. If not, the system will take the earlier computed value of the electron charge density as the starting value and re-start over again. This process will be continued until convergence has been achieved.

The SCF cycle is perfectly employed in the PWSCF code of the QUANTUM ESPRESSO simulation suite. In this research, this software package is used to determine the electronic properties of the two heterobilayers, namely Sn/SiC and Plumbene/h-BN.

### 3.5.2 DFT Method in the Study of Heterobilayer

PWSCF code implemented in the QUANTUM ESPRESSO is used to accomplish the electronic and ionic computation by implementing the program PW.x [60] of QUANTUM ESPRESSO. This program runs the SCF cycle to calculate the ground state energy. In PW.x, the input files are written in a text editor where the system receives a command to complete the determination of convergence.To calculate the energy bandgap and band structure, QUANTUM ESPRESSO needs three programs: pw.x, bands.x, and plotband.x [60]. The NAMELISTS and INPUT_CARDS make the pw.x input files. Therefore, three input formats are compulsory in the NAMELISTS:

- o &CONTROL defines the type of calculation to be performed and the quantity of inputs/outputs.
- o &SYSTEM these are input variables that specify a particular system. &ELECTRONS the variables that are written here decide which particular algorithm will be used to have a self-consistent Kohn-Sham equations' solution.

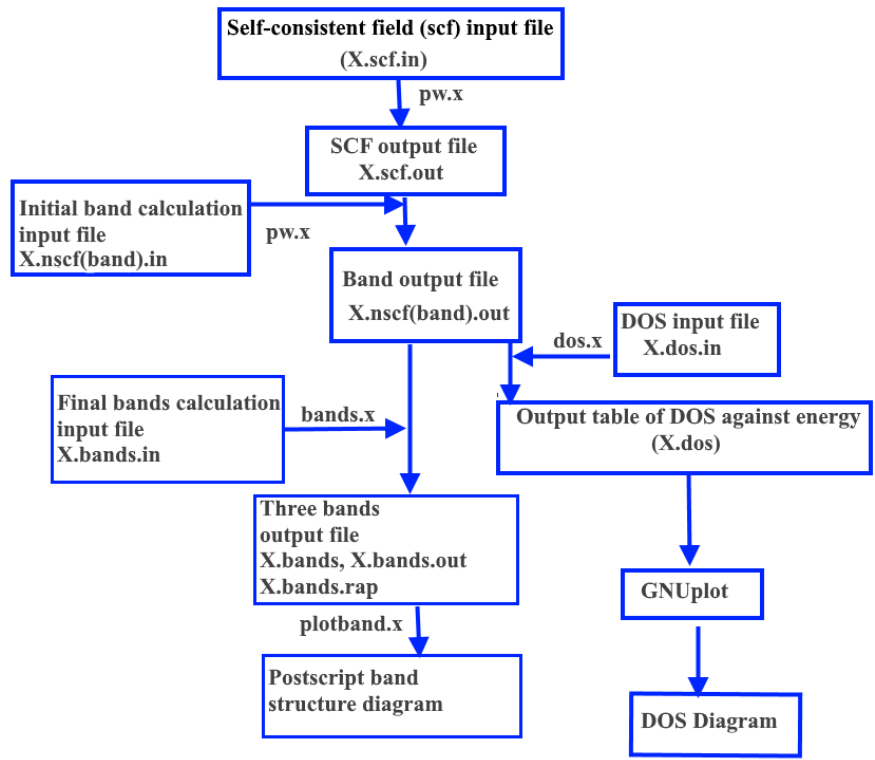

Figure 3.5: Procedures to calculate the electronic band diagram and density of states with QUANTUM ESPRESSO

The INPUT_CARDS contain three compulsory files:

- o ATOMIC_SPECIES conveys information on the name, mass, and pseudo-potentials of the elements used in the system.
- o ATOMIC_POSITIONS carry the information about the coordinates of every atom of the primitive unit cell.
- o K-POINTS specifies the coordinates as well as weights of the **k**-points taken for the Brillouin zones

The procedures required to calculate the energy band diagram and density of states (DOS) in the QUANTUM ESPRESSO simulation package are illustrated in the block diagram of Figure 3.5.

Three input files namely (i) GeAlP.scf.in (ii) GeAlP.b-nscf.in and (iii) GeAlP.band.in which are used to calculate the electronic band structure of Sn/SiC heterobilayer are shown below:

### **GeAlP.scf.in**

```
&CONTROL
calculation          = "scf"
outdir               = "./work/"
prefix               = "GaP+Ge_I_3.70"
pseudo_dir           = "./pseudo/"
restart_mode         = "from_scratch"
verbosity            = 'high'
/
&SYSTEM
ibrav                = 4
a                    = 3.89
c                    = 20
nat                  = 4
ntyp                 = 3
ecutwfc              = 30.0,
ecutrho              = 120.0,
occupations          = 'smearing'
smearing             = 'm-p'
degauss              = 0.0005
input_dft            = 'pbe'
vdw_corr             = 'Grimme-D2',
/
&ELECTRONS
```

conv_thr                = 1.00000e-8
mixing_beta             = 0.7
/
ATOMIC_SPECIES
Ga 69.723      Ga.pbe-mt_fhi.UPF
P 30.973762 P.pbe-mt_fhi.UPF
Ge 72.64 Ge.pbe-mt_fhi.UPF

ATOMIC_POSITIONS (angstrom)
Ga     0.000000000  0.000000000 0.000000
P      0.000000000  2.245892547 0.380000
Ge     0.000000000  0.000000000  3.70
Ge     0.000000000  2.245892547 4.08

K_POINTS {automatic}
10  10  1  0  0  0

## **GeAlP.b-nscf.in**

&CONTROL
calculation             = "bands"
outdir                  = "./work/"
prefix                  = "GaP+Ge_I_3.70"
pseudo_dir              = "./pseudo/"
restart_mode            = "from_scratch"
verbosity               = 'high'
/
&SYSTEM
ibrav                   = 4
a                       = 3.89
c                       = 20
nat                     = 4
ntyp                    = 3
ecutwfc                 = 30.0,
ecutrho                 = 120.0,
occupations             = 'smearing'
smearing                = 'm-p'
degauss                 = 0.0005
input_dft               = 'pbe'
vdw_corr                = 'Grimme-D2',
/
&ELECTRONS
conv_thr                = 1.00000e-8
mixing_beta             = 0.7
/
ATOMIC_SPECIES

```
Ga 69.723      Ga.pbe-mt_fhi.UPF
P 30.973762 P.pbe-mt_fhi.UPF
Ge 72.64 Ge.pbe-mt_fhi.UPF

ATOMIC_POSITIONS (angstrom)
Ga      0.000000000  0.000000000 0.000000
P       0.000000000  2.245892547 0.380000
Ge      0.000000000  0.000000000 3.70
Ge      0.000000000  2.245892547 4.08

K_POINTS {crystal_b}
4
0.0000000 0.0000000 0.0000000 20 !G
0.3333333 0.3333333 0.0000000 20 !k
0.0000000 0.5000000 0.0000000 20 !M
0.0000000 0.0000000 0.0000000 20 !G
```

**GeAlP.band.in**

```
&bands
outdir              = "./work/"
prefix              = 'GaP+Ge_I_3.70'
filband             = 'GaP+Ge_I_3.70.band'
lsym                = .true.
/
```

## 3.6   XCRYSDEN: Visualization Tool

A visualization package, XCRYSDEN [60], is used to visualize the crystal structure that was implemented in QUANTUM ESPRESSO. Very good knowledge of the atomic points of the materials being examined is mandatory because the SCF input file uses that data. Using XCRYSDEN, one can see the crystal structure pattern and check it aggress with the desired structure. For that, it is necessary to run the SCF input file with XCRYSDEN. The process is shown in Figure 3.6

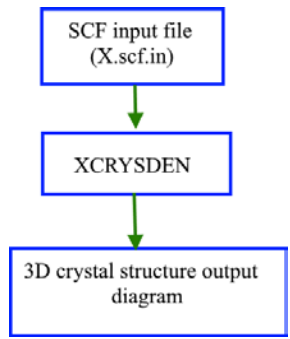

Some properties of XcrySDen should be mentioned with relative functions:

- XCrySDen as a Molecular visualization program.
- XCrySDen as a property analyzer.
- Plotting Properties with the 2D Graphing System.
- 3D Property Visualizer.
- Displaying the Properties on a Selected Plane.

### 3.7 Method of Calculation

Figure 3.6: Procedure to use XCRYSDEN to see the crystal structure

### 3.7.1 Germanene/2d-GaP Heterobilayer

For the first principle investigation of the germanene/GaP heterostructure, the DFT framework has been considered. The conventional Quantum Espresso [60] environment is taken to simulate the model and analyze the outcomes. Troullier-Martin's norm-conserving pseudopotential has been used to represent the electron-ion interactions [63] and to account for the spin-orbital coupling. The exchange and correlation interaction is reckoned by taking the Perdew-Burke-Ernzerhof (PBE) functional under the generalized gradient approximation (GGA) [64]. The cut-off kinetic energy of the plane wave function is set at 30 Ry. Methfessel-Paxton smearing is utilized for initiating smearing occupation. A sampling of the first Brillouin zone is done by taking a $10 \times 10 \times 1$ Monkhorst- Pack grid scheme [65]. Further, a $22 \times 22 \times 1$ Monkhorst-Pack grid scheme is to be taken for ensuring proper sampling whilst calculating the density of states and partial density of states. The structure is to be well relaxed, meeting an energy convergence of the order of $1e^{-8}$ and ensuring self- consistency. Semi-empirical Grimme DFT-D2 [66] correlation has been considered to evaluate the van der Waals interaction throughout the computation. Again a vacuum space of about 20 Å is proposed in between the bilayer to prevent interaction between them and with the vicinity normal to the single individual layers. To demonstrate the properties of Ge/GaP heterobilayer, three stacking patterns are to be considered. These patterns are then to be well relaxed to converge in the cutoff limits. The binding energy between Ge and SiC layer can be calculated by

$$E_b = \left( E_{Ge/GaP} - E_{Ge} - E_{GaP} \right)/A$$

In addition to that, the binding energy per Ge atom can be shown as

$$E_{Ge} = \left( E_{Ge/GaP} - E_{Ge} - E_{GaP} \right)/N_{Ge}$$

Where $E_{Ge/GaP}$ represents the energy of the heterostructure, $E_{Ge}$ and $E_{GaP}$ are the layer energy amplitude; $A$ stands for the area covered and $N_{Ge}$ is the number of Ge atoms in the optimized structure. Our optimized structure provides Ge with a lattice constant of 3.95 Å with buckling length 0.69 Å, which is in closer approximation with the research conducted previously by Garcia JC *et al.* [22] and Roome *et al.* [67] & Shyam *et al.* [56]. We got a lattice constant of 3.89 Å with buckling length 0.38 Å for the substrate GaP to make the heterobilayer; this is in close proximity to the result found by V. Ongun Ozc., *et al.* [30]. Since these lattice constants are not the same, a mismatch of about 1.5% is considered with applying strain on the Ge layer. We utilized a 1×1 Ge layer with a 20 Å of a vacuum above the 1×1 GaP layer.

### 3.7.2 Germanene/2d-AlP Heterobilayer

The similar atomic variable mentioned above is also utilized in germanene/AlP heterostructure; the binding energy equation resembles the same with the replacement of Ga atoms by Al atoms. Troullier-Martin [63] norm-conserving pseudopotential, Perdew-Burke-Ernzerhof (PBE) functional under the generalized gradient approximation (GGA) [64] is utilized. A sampling of the first Brillouin zone is done by taking a 10×10×1 Monkhorst-Pack grid scheme[65]. Further, a 22×22×1 Monkhorst-Pack grid scheme is to be taken for ensuring proper sampling whilst calculating the density of states and partial density of states. The optimized lattice constant obtained for Al is 3.95 Å, the stable 2D Al layer is essentially a planar structure. This causes to take a 1×1 Ge layer above the 1×1 AlP layer considering a mismatch amount of almost 0.01%. The optimized lattice constant obtained is in closer proximity to the findings of V. Ongun Ozc. *et al.* [30] shown in their investigations are to be used for this research work.

# CHAPTER IV

## Results and Discussions

### 4.1 Introduction

This chapter accumulates the results that were obtained while investigating with two Nobel heterostructures naming: germanene/2d-GaP and germanene/2d-AlP. In the DFT framework, simulation was operated on the Quantum Espresso suite [60]. The overall electronic properties gathered here will then be used to predict the heterostructures duos applicability, and in the last chapter, concluding remarks will wrap up the thesis work.

### 4.2 Tunable Electronic Properties of Germanene/2d-GaP Heterostructure

Germanene/2d-GaP is our first material of concern. Planar hexagonal germanene is placed on top of the hexagonal GaP monolayer in 2d. Different electronics properties are going to be investigated in the following research contents.

### 4.2.1   Structures and Binding Energy calculation of Germanene/2d-GaP Heterostructure

Ge/2d-GaP heterostructure is considered for investigating different properties. The supercell of Ge/2d-GaP constitutes four atoms covering 2 Ge atoms over 2 GaP atoms. This pattern consideration is comparable to the heterostructure consideration of Sn/h-BN by Khan *et al.* [60], Sn/SiC by Naim *et al.* [39]. Our considered 1.5% lattice mismatch matches closely to the study of Asmaul *et al.* with graphene/SiC structure [68], Ren *et al.* with stanene/MoS₂ [68]. Our optimized unit cells of Germanene and 2d-GaP have been shown in Figure 4.1. The figure describes the zero bandgaps on the 2D Ge structure at the K point. AlP is a wide bandgap material having a bandgap of the magnitude of 1.53 eV [30]. However, our optimized structure with norm-conserving pseudo-Potential shows that about 1.73 eV bandgap is actually obtained from the 2-d low buckled GaP unit cell. Our result is close to the data obtained by V. Ongun Ozc. and co-workers while working for different 2d semiconductor substrates with PBE exchange-correlation [30]. Moreover, similar observation results during the investigation of 2D germanene and silicene with SiC substrate. The single germanene layer for the mentioned 2D material got zero bandgaps when are retained isolated from the GaP layer. Three stacking patterns are being considered within this research work, as depicted in Figure 4.2; the atomic layout is also shown.

The three stacking patterns that have been undertaken are depicted in figure 4.2. The three structural patterns are formed by placing armchair edge Ge atom over the Ga atom, over the P atom, and over the Ga atom with a shift in position, respectively. Another way to define is to change the armchair edge position of the buckled GaP substrate with Ga for pattern-I, P for pattern-II, and Ga with a shift from pattern-I position forming pattern-III.

The relax calculation provides the convergence of the binding energy of structures I, II, and III being at the interlayer displacement of 3.70 Å, 2.96 Å & 3.52 Å, respectively, as shown in Figure 4.3. It was observed that when the phosphorus atom is placed just below the Ge atom at the edge of buckled germanene, the higher binding energy per germanene atom results. The binding energy level shows the bonding strength between the layers. Greater negative value shows higher stability of the patterns. Binding energy less than -180 meV refers to the weak Van der Waals bonding. The binding energy and binding energy per Ge atom for the structures are tabulated in Table 4.1. The tabulated data illustrates the most stable positions of each structure.

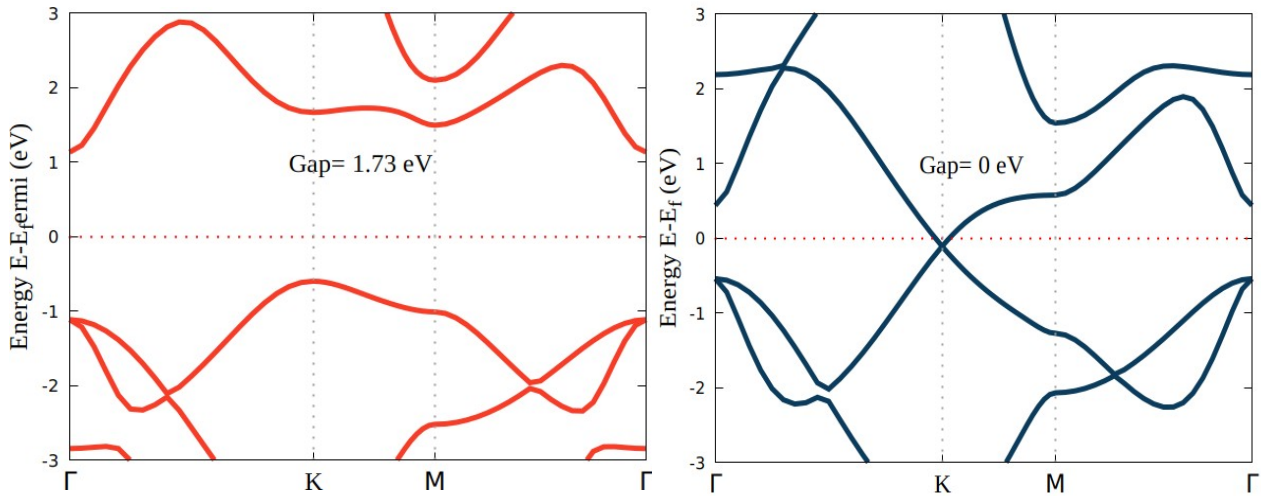

Figure 4.1: The unit cell band structure of (a) Gallium Phosphide, GaP. (b) Germanene, Ge.

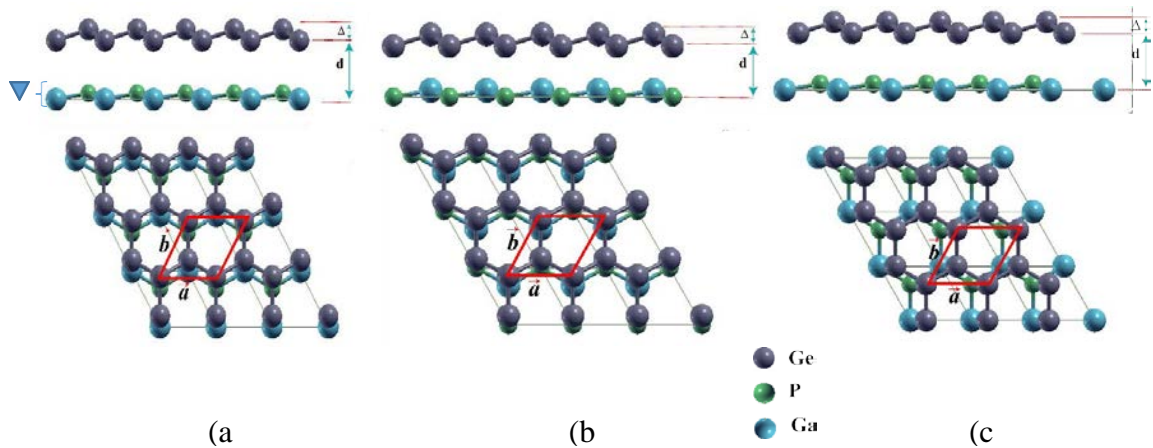

Figure 4.2: Three representational stacking configurations of Ge/GaP heterobilayer at equilibrium (a) Pattern-I (b) Pattern-II (c) Pattern-III. 'd' denotes the inter-layer distance between Ge and GaP monolayer, and 'Δ' is the buckling height.

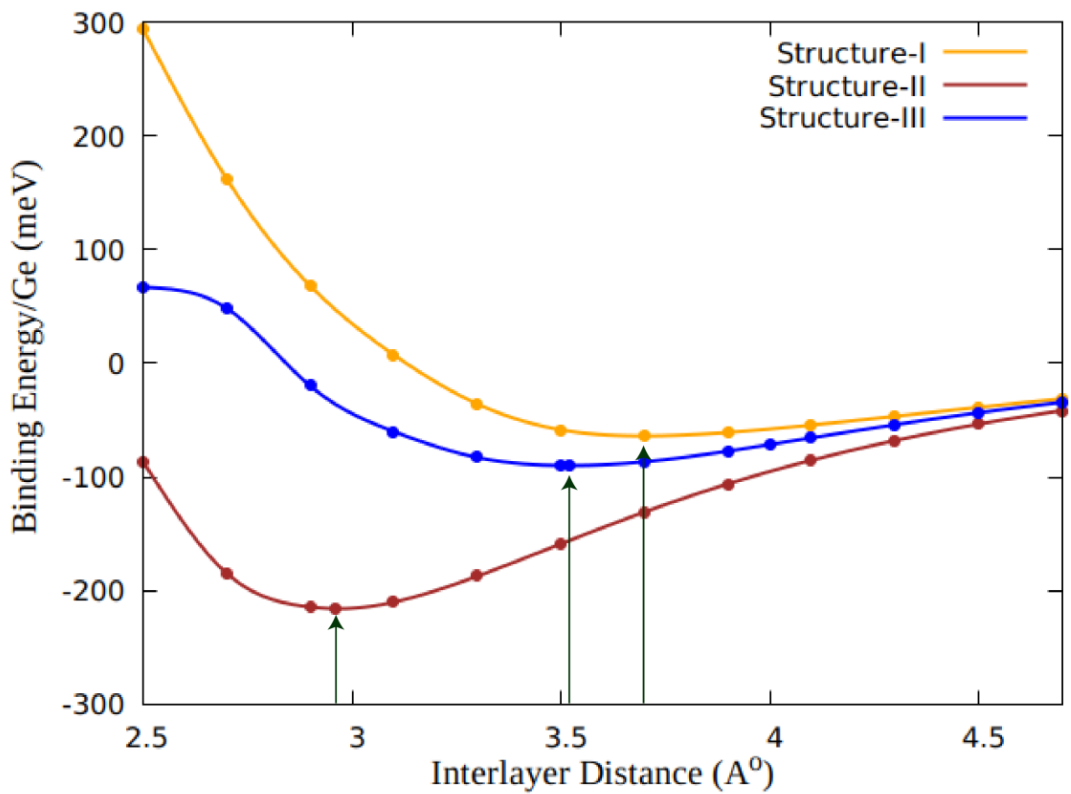

Figure 4.3: The variation of binding energy per germanene atom as a function of interlayer distance for three stacking configurations of Ge/GaP heterobilayer. Up arrows indicate the optimized interlayer distances.

Our observation of the three patterns shows that the presence of weak bonding exists between the layers. Binding energy per Sn atom during the work of Sn/SiC heterobilayer resulted in around -230 meV [39]. While investigating further binding energy found for graphene/SiC resulted in about -320 meV [68].

Table 4.1: Structural and electronic properties of Ge/GaP heterobilayer for the three configurations, comprising binding energies per Sn atoms (Eb), bandgap at K point (Eg) using LDA and HSE functional, optimized interlayer distance (d).

| Configuration | Eb/Ge atom (meV) | Eg (meV) with LDA* | Eg (meV) with HSE* | d (Å) |
|---------------|------------------|--------------------|--------------------|-------|
| Structure-I   | -63.9741         | 162.2              | -127.9482          | 3.70  |
| Structure-II  | -215.853         | 154.7              | -431.706           | 2.96  |
| Structure-III | -90.023          | 159.8              | -180.046           | 3.52  |

This information improvises the acceptability of our optimized heterobilayer structures. This binding energy is important to realize the ease for structural realization and for applicability. A hexagonal structure with about 0.69 Å buckling of Ge configuration is configured for each pattern. Similar structural configurations were taken by Naim *et al.*, Khan *et al.* while working with Sn/SiC and Sn/h-BN, respectively [33,39]. Buckled structures are often treated as responsible for broadening a direct bandgap on the high symmetrical points of the first Brillouin zone. However, the influence on the germanene atoms differs as the spatial positions of GaP atoms change in each pattern.

### 4.2.2 Band Diagram and DOS of Germanene/2d-GaP Heterobilayer

The interaction of the bands of the Ge, Ga, and P are to be investigated to have an insight into the overall band formation and band properties. To have an insight into the electron-ion interactions and other classical properties, the band diagram, along with the respective density of states, is shown in Figure 4.4. We observed an opening of the bandgap in the 2D Ge layer due to interaction with huge bandgap GaP. The band diagram shows the existence of a bandgap of about 463.2 meV, 285.1 meV, 154.4 meV for the pattern-I, II & III, respectively. The bandgap achieved during graphene/SiC was about 25 meV, 20 meV, and 28 meV for pattern-I, pattern- II & pattern-III, respectively [68]. Thus our results converge well with the ongoing approach to broadening bandgap. This gap opening in the direct cones paves the way to use Ge in nanoelectronics. Moreover, the band diagram gives the clear idea is that rather than the wide bandgap GaP, the Ge layer mainly demonstrates the electronic structure since it mainly contributes to the bandgap formation. Huge band gap opening exhibits the break-in germanene lattice symmetry; thus, GaP substantially influences the Ge monolayer properties. The use of relativistic pseudopotential to describe the spin-orbital coupling (SOC) is also investigated. Spin orbital may increase or decrease the bandgap depending upon the form of interaction between the atoms. 2D Sn/SiC, C/SiC, C/h-BN were also characterized with change in bandgap after SOC association [33,68]. Figure 4.5 shows the band diagram of our considered three patterns. The bandgap decreases for each structure and becomes 456.5 meV, 294.3 meV, 151.4 meV, respectively. To characterize the inner properties, the energy states are also attached to Figure 4.5. The density of state depicts the association of Ge and GaP bands to evolve a band. Tetrahedra occupation method with GGA-PBE approach provides states such that above and below the Fermi level at gamma point the Ge band are dominating, this illustrates Ge to be the responsible one to demonstrate the hetero-structure characteristics.

### 4.2.3 PDOS and CDD of Germanene/2d-GaP Heterostructure

A closer look can be achieved by looking at Figure 4.6, which shows the partial density of the state. The contribution of each of the atomic orbitals has also been plotted. From the PDOS plot, it is observed that the p-orbital of germanene plays a major part. The state in the Ge-p spreads highly in the conduction band, proving that germanene carries the carrier. Thus the inherent properties of germane may be achieved while utilizing the heterostructure. Furthermore, P-p and Ga-p orbitals also influence the valence and conduction bands formation. Therefore, the density of states will exhibit the germanene properties mostly.

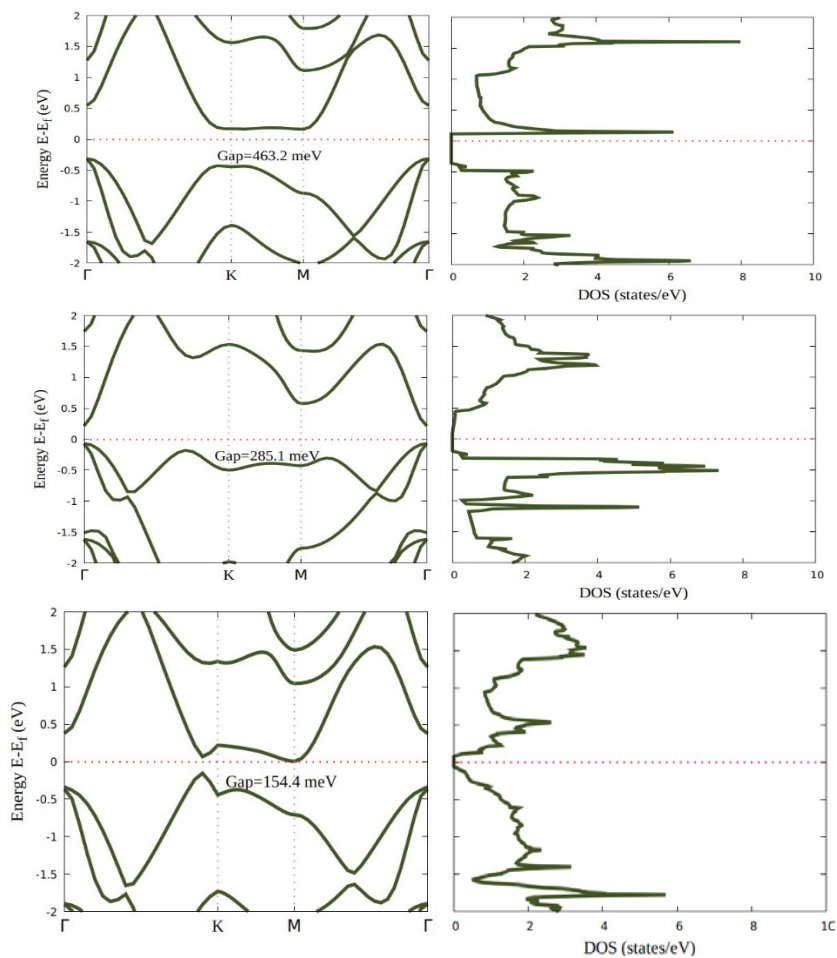

Figure 4.4: Band diagrams and the corresponding density of states (DOS) of the Ge/GaP heterobilayers for (a) structure-I (b) structure-II and (c) structure-III.

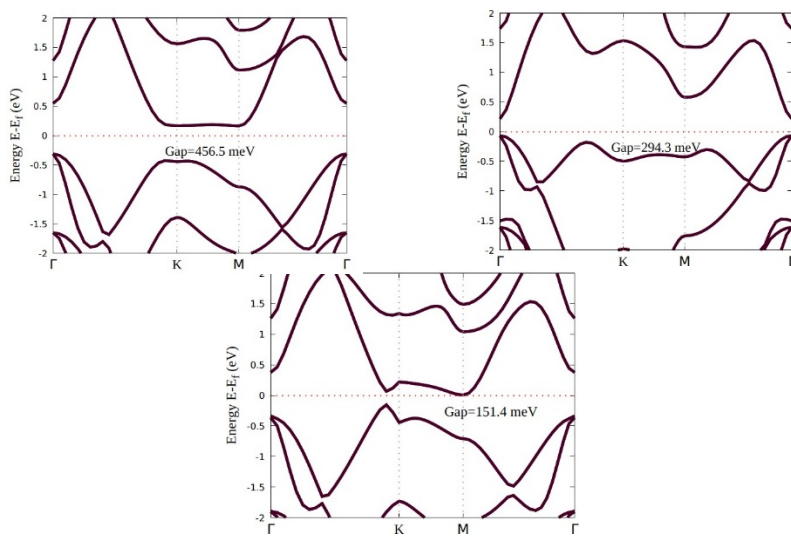

Figure 4.5. Band structures of the Ge/GaP heterobilayer, including SOC of (a) structure-I (b) structure-II and (c) structure-III.

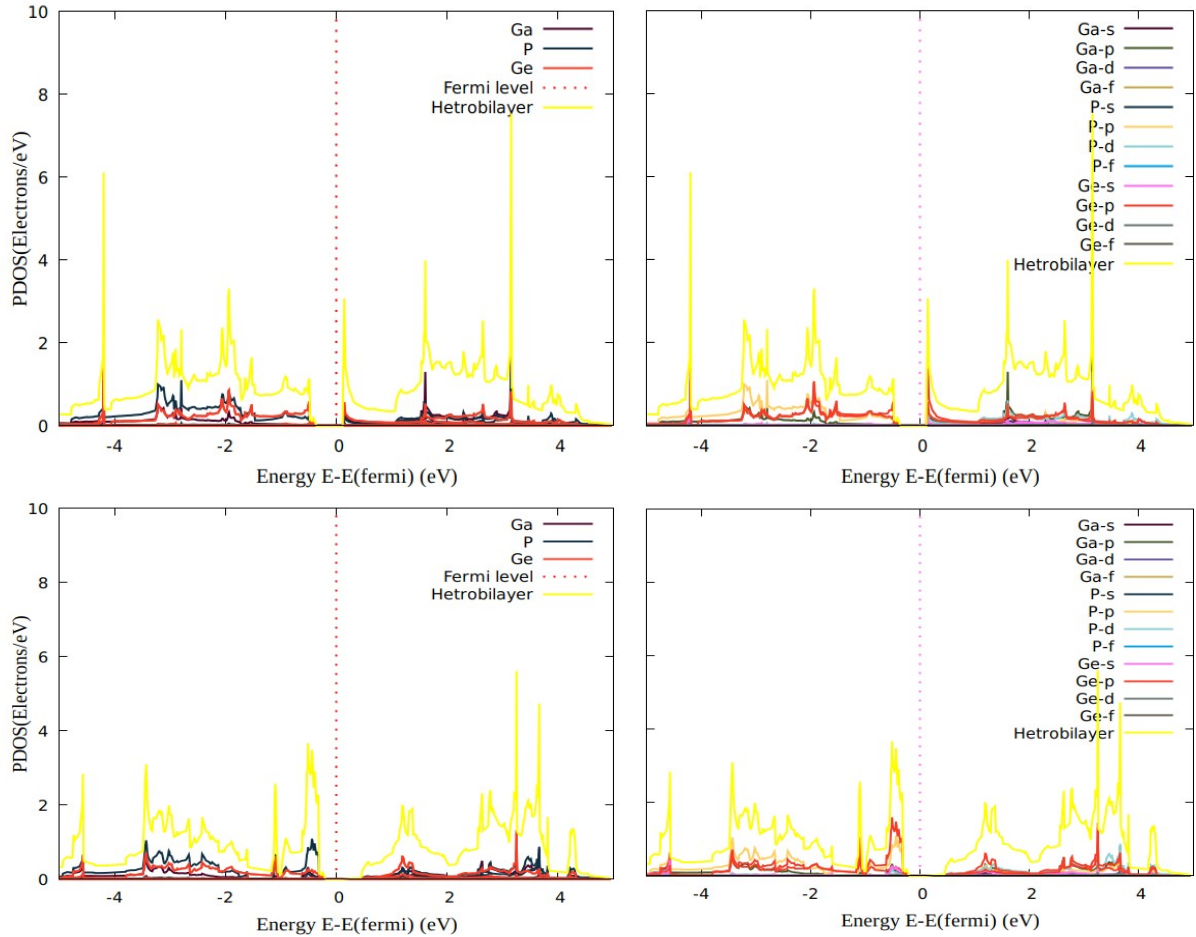

Figure 4.6: Atom projected density of states for structure-I and structure-II of the Ge/GaP heterobilayer. The Fermi level is made 0. Overall contribution from Ga, P, and Ge atoms of (a) Structure-I and (b) Structure-II are exhibited.

The influence of Ge on the overall structure can further be demonstrated with Figure 4.7, which shows the charge density distribution for pattern-I. This clearly shows that the p-orbital of Ge is to be spread out, thus motivating the carrier concentration through the Ge layer only. Since GaP doesn't contribute much to carrier transportation, this also illustrates the expediency of choosing GaP as the substrate for the structure. Since all the structure resembles about the same distributions, we utilized pattern-II for the calculation of space charge density distribution and CDD.

From the charge density difference as shown in Figure 4.7 for pattern-I, the contribution of germanene at the space charge layer is clearly depicted. The charge contribution within the germanene layer changes, providing the breaking in symmetry. Moreover, electron transfers from the germanene layer to the GaP layer occur. This causes the evolve in a bandgap. Since most of the charge is convoluted in the bucked germanene structure, carrier mobility is controlled only by the germanene layer. This phenomenon observes well-matched with the work done during the investigation of C/SiC, Sn/SiC, C/MoS$_2$, C/h-BN 2D heterostructures [68,33]. Thus the figure gives insight into the resultant band gap and energy state.

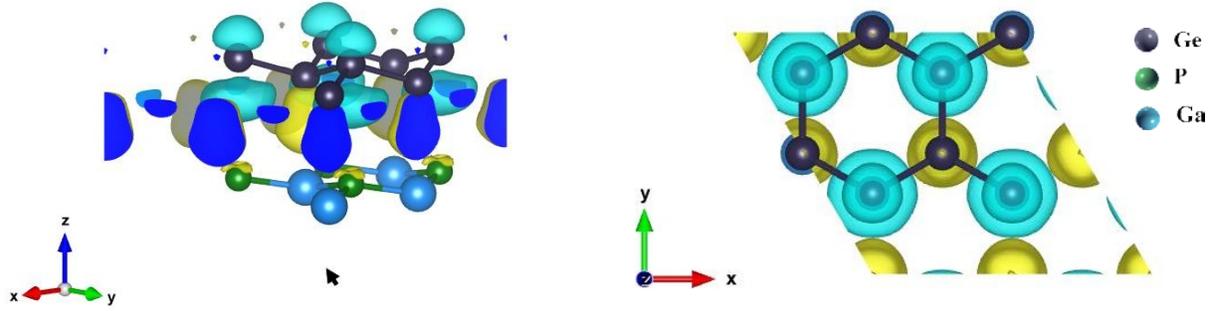

Figure 4.7: The charge density difference of the Ge/GaP heterobilayer for Structure-I. Cyan and Yellow indicate the increase and decrease of charge in the space regarding free standing germanene and GaP layer, respectively. The isovalue is 0.004 e/Å3. (a)

Due to charge transfer within the germanene sublattices, the overall symmetry breaks down. Overlap of the layers causes the localizations of charges which may be the direct effect of orbital overlap or gradient of potential from layer on the close vicinity or may be due to polarization.

### 4.2.4 Band Gap Tuning with Variation of Interlayer Distance and Bi-Axial Strain

The germanene/GaP heterobilayer can be analyzed further for its bandgap tuning properties with respect to the interlayer distances, as demonstrated in Figure 4.9. With the increase in an interlayer distance, fewer interactions occur between the layers, less charge overlap & crystal symmetry distortion occurs. As a result of this phenomenon, less bandgap will induce at the Dirac cones. Again, if we reduce the interlayer distance from the convergence value, the band gap decreases, but this time, the band overlaps to a higher magnitude to commensurate the tight-binding model and modern physics. The tenability of the bandgap in the heterostructure can be utilized for making spintronic devices quite efficiently as it is seen that the pattern-I & pattern-II follow similar characteristics throughout the investigation. The pattern-I with a slightly sluggish change in binding energy and band gap variation with interlayer distance in the range < 3.1 Å exhibits the

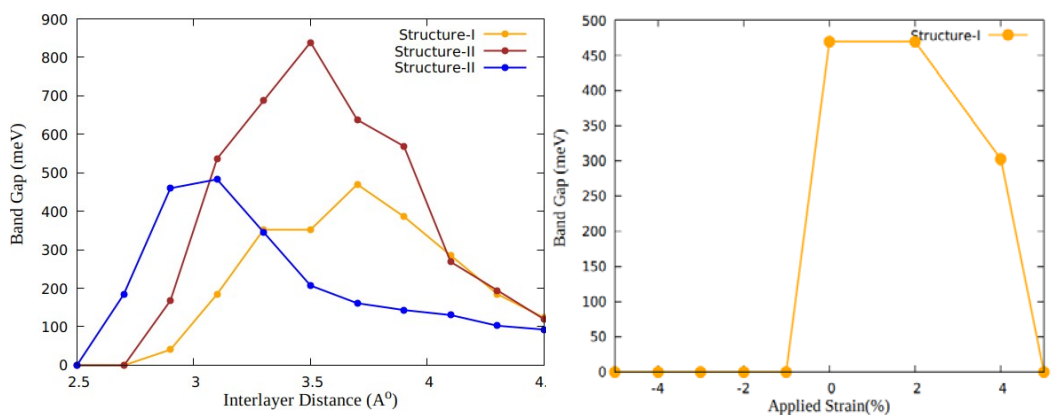

Figure 4.8: Band tuning of Structure-I with variation in (a) interlayer distance, (b) applied bi-axial tensile strength. Negative and positive strain refers to tensile and compressive strain.

change of the former two with a slightly lower amplitude.The bandgap tuning may also be achieved with a change in applied strain to the germanene layer, as shown in Figure 4.8. It is expected that the bandgap decreases slightly with an increase in strain associated with a significant fermi level, as observed by Asmaul *et al.* during the investigation of graphene/SiC heterostructure [68]. And an opposite change might be observed for compressive strain. More applied compressive or repulsive strain can cause the break in lattice symmetry and introduce variations in band gap at the Brillouin zone symmetrical points. From table 4.2, it's clearly depicted that with an increase in interlayer distance, the bandgap is changeable and a shift from indirect to direct bandgap at K-point at 1st BZ is observable. Below 3.3 Å, an indirect bandgap occurs

Table 4.2: Level of bandgap, its position, and type with an increase in interlayer displacement d(Å)

| Interlayer distance, d(Å) | Band Gap (meV) | Position of Band Gap (1st BZ) | Type of Band Gap |
|---|---|---|---|
| 2.5 | 0 | M+(G—K) | Indirect |
| 2.7 | 0 | M+(G--K) | Indirect |
| 2.9 | 40.3 | M+(G--K) | Indirect |
| 3.1 | 184.6 | M+(G--K) | Indirect |
| 3.3 | 352.4 | M-G | Indirect |
| 3.5 | 352.4 | M-G | Indirect |
| 3.7 | 469.8 | M-G | Indirect |
| 3.9 | 386.0 | K | Direct |
| 4.1 | 285.3 | K | Direct |
| 4.3 | 184.6 | K | Direct |
| 4.5 | 123.2 | K | Direct |

Between M point and a point between Gamma and K-point that shifts towards the K point with an increase in d. At 3.3, an indirect bandgap occurs between M and Gamma–symmetrical points, which with further increase in 'd' results in a direct bandgap at K-point.

## 4.3 Tunable Electronic Properties of Germanene/2d-AlP Heterostructure

Germanene/2d-AlP is our first material of concern. Planar hexagonal germanene is placed on top of the hexagonal AlP monolayer in 2d. Different electronics properties are going to be investigated in the following research contents.

### 4.3.1   Structures and Binding Energy calculation of Germanene/2d-AlP

Ge/2d-AlP heterostructure is considered for investigating different properties. The supercell of Ge/2d-AlP constitutes four atoms covering 2 Ge atoms over 2 AlP atoms. This pattern consideration is comparable to the heterostructure consideration of Sn/h-BN by Khan *et al.* [60], Sn/SiC by Naim *et al.* [39]. Our considered 1.5% lattice mismatch matches closely to the study of Asmaul *et al.* with graphene/SiC structure [68], Ren *et al.* with stanene/MoS₂ [68]. Our optimized unit cells of Germanene and 2d-AlP have been shown in Figure 4.9. The figure describes the zero bandgaps on the 2D Ge structure at the K point. AlP is a wide bandgap material having a bandgap of the magnitude of 1.53 eV [30]. However, our optimized structure with the norm-conserving pseudopotential shows that about 2.25 eV bandgap is actually obtained from the 2-d low buckled GaP unit cell. Our result is close to the data obtained by V. Ongun Ozc. and co-workers while working for different 2d semiconductor substrates with PBE exchange-correlation [30]. Moreover, similar observation results during the 2D germanene and silicene investigation with SiC substrate. The single germanene layer for the mentioned 2D material got zero bandgaps when are retained isolated from the AlP layer. Three stacking patterns are being considered within this research work, as depicted in Figure 4.10; the atomic layout is also shown.

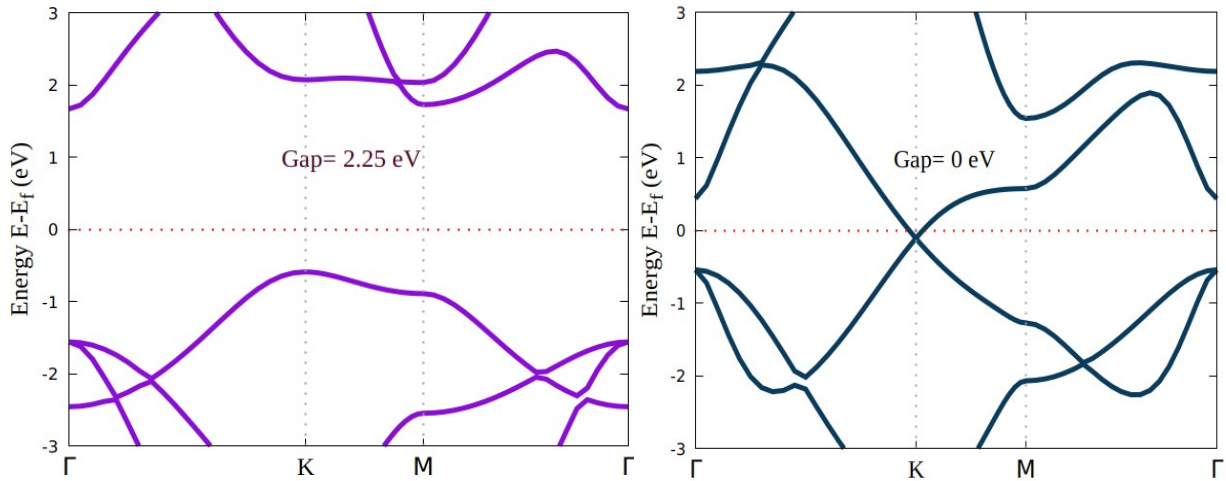

Figure 4.9: The unit cell band structure of (a) Aluminium Phosphide, AlP. (b) Germanene, Ge.

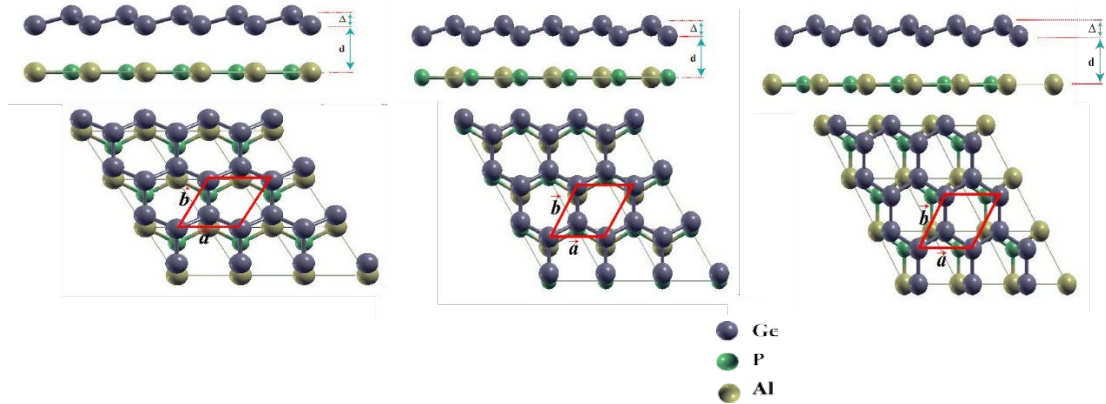

Figure 4.10: Three representational stacking configurations of Ge/AlP heterobilayer at equilibrium (a) Pattern-I (b) Pattern-II (c) Pattern-III. 'd' denotes the inter-layer distance between Ge and GaP monolayer, and 'Δ' is the buckling height of the germ

The three stacking patterns that have been undertaken are depicted in figure 4.10. The three structural patterns are formed by placing the armchair edge Ge atom over the Al atom, the P atom, and the Al atom with a shift in position, respectively. Another way to define is to change the armchair edge position of the buckled AlP substrate with Al for pattern-I, P for the pattern- II, and Al with a shift from pattern-I position forming pattern-III.

The relax calculation provides the convergence of the binding energy of structures I, II, and III being at the interlayer displacement of 3.24 Å, 2.71 Å, and 3.13 Å, respectively, as shown in Figure 4.11. It was observed that when the phosphorus atom is placed just below the Ge atom at the edge of buckled germanene, the higher binding energy per germanene atom results. The binding energy level shows the bonding strength between the layers. Greater negative value shows higher stability of the patterns. Binding energy less than -180 meV refers to the weak Van der Waals bonding. The binding energy and binding energy per Ge atom for the structures are tabulated in Table 4.2. The tabulated data illustrates the most stable positions of each structure. Our observation of the three patterns shows that the presence of weak bonding exists between the layers. Binding energy per Sn atom during the work of Sn/SiC heterobilayer resulted in around -230 meV [39]. While investigating further binding energy found for graphene/SiC resulted in about -320 meV [68].

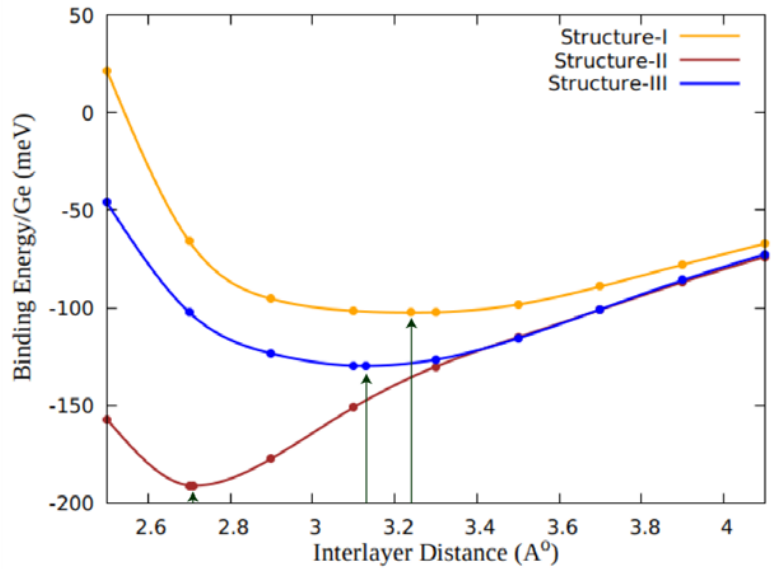

Figure 4.11: The variation of binding energy per germanene atom as a function of interlayer distance for three stacking configurations of Ge/AlP heterobilayer. Down arrows indicate the optimized interlayer distances.

Table 4.3: Structural and electronic properties of Ge/AlP heterobilayer for the three configurations, comprising binding energies per Sn atoms (Eb), bandgap at K point (Eg) using LDA and HSE functional, optimized interlayer distance (d)

| Configuration | Eb/Ge atom (meV) | Eg (meV) with LDA* | Eg (meV) with HSE* | d (Å) |
|---|---|---|---|---|
| Structure-I | -102.454 | 162.2 | -240.908 | 3.24 |
| Structure-II | -190.985 | 154.7 | -381.97 | 2.71 |
| Structure-III | -129.677 | 159.8 | -259.354 | 3.13 |

This information improvises the acceptability of our optimized heterobilayer structures. This binding energy is important to realize the ease for structural realization and for applicability. A hexagonal structure with about 0.69 Å buckling of Ge configuration is configured for each pattern.

Similar structural configurations were taken by Naim *et al.*, Khan *et al.* while working with Sn/SiC and Sn/h-BN, respectively [33,39]. Buckled structures are often treated as responsible for broadening a direct bandgap on the high symmetrical points of the first Brillouin zone. The influence on the germanene atoms differs as the spatial positions of AlP atoms change in each pattern.

### 4.3.2    Band Diagram and DOS of Germanene/2d-AlP Heterobilayer

The interaction of the bands of the Ge, Al, and P are to be investigated to have an insight into the overall band formation and band properties. To have an insight into the electron-ion interactions and other classical properties, the band diagram, along with the respective density of states, is shown in Figure 4.4. We observed an opening of the band gap in the 2D Ge layer due to interaction with huge bandgap AlP. The band diagram shows the existence of a band gap of about 490.0 meV, 290.3 meV, 264.3 meV for the pattern-I, II & III, respectively. The bandgap achieved during graphene/SiC was about 25 meV, 20 meV, and 28 meV for pattern-I, pattern- II & pattern-III, respectively [68]. Thus our results converge well with the ongoing approach to broadening bandgap. This gap opening in the direct cones paves the way to use Ge in nanoelectronics. Moreover, the band diagram gives the clear idea is that the Ge layer mainly demonstrates the electronic structure rather than the wide bandgap ALP since it mainly contributes to the bandgap formation. Huge band gap opening exhibits the break-in germanene lattice symmetry; thus, AlP substantially influences the Ge monolayer properties. The use of relativistic pseudopotential to describe the spin-orbital coupling (SOC) is also investigated. Spin orbital may increase or decrease the bandgap depending upon the form of interaction between the atoms. 2D Sn/SiC, C/SiC, C/h-BN were also characterized with change in bandgap after SOC association [33,68]. Figure 4.12 shows the band diagram of our considered three patterns. The bandgap decreases for each structure and becomes 483.3 meV, 287.8 meV, 270.2 meV, respectively. To characterize the inner properties, the energy states are also attached within Figure 4.13. The density of state depicts the association of Ge and AlP bands to evolve a band. Tetrahedra occupation method with GGA-PBE approach provides states such that above and below the Fermi level at gamma point the Ge band are dominating, this illustrates Ge to be the responsible one to demonstrate the hetero-structure characteristics.

### 4.3.3    PDOS and CDD of Germanene/2d-AlP Heterostructure

A closer look can be achieved by looking at Figure 4.14, which shows the partial density of the state. The contribution of each of the atomic orbitals is also plotted. From the PDOS plot, it is observed that the p-orbital of germanene plays a major part. The state in the Ge-p spreads highly in the conduction band, proving that germanene carries the carrier. Thus the inherent properties of germane may be achieved while utilizing the heterostructure. Furthermore, P-p and Al-p orbitals also influence the valence and conduction bands formation. Therefore, the density of states will exhibit the germanene properties mostly.

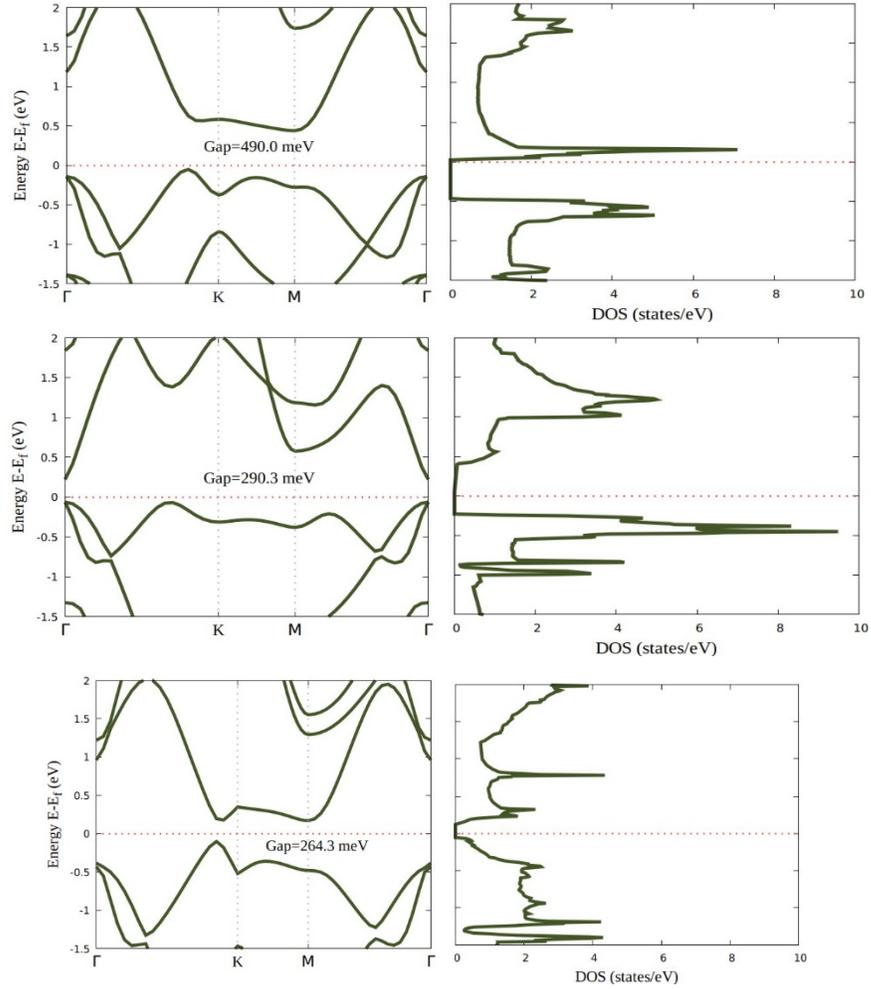

Figure 4.12: Band diagrams and the corresponding density of states (DOS) of the Ge/AlP heterobilayers for (a) structure-I (b) structure-II and (c) structure-III.

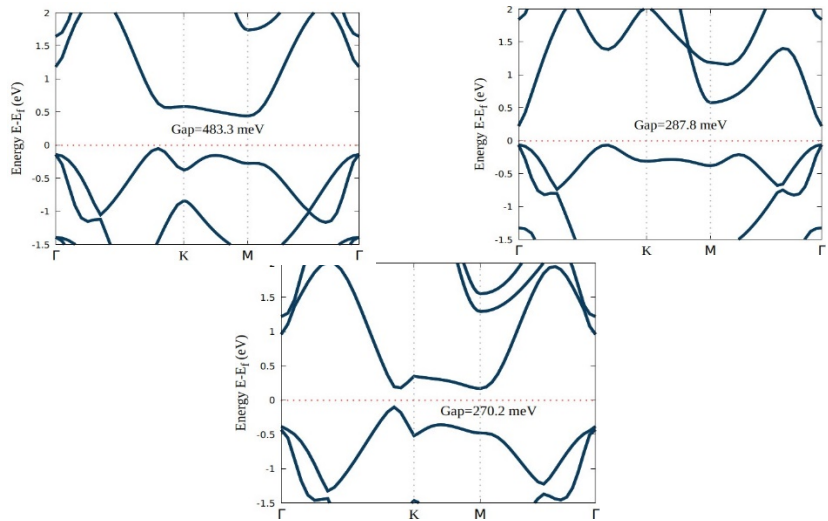

Figure 4.13: Band structures of the Ge/AlP heterobilayer including SOC of (a) structure-I (b) structure-II and (c) structure-III.

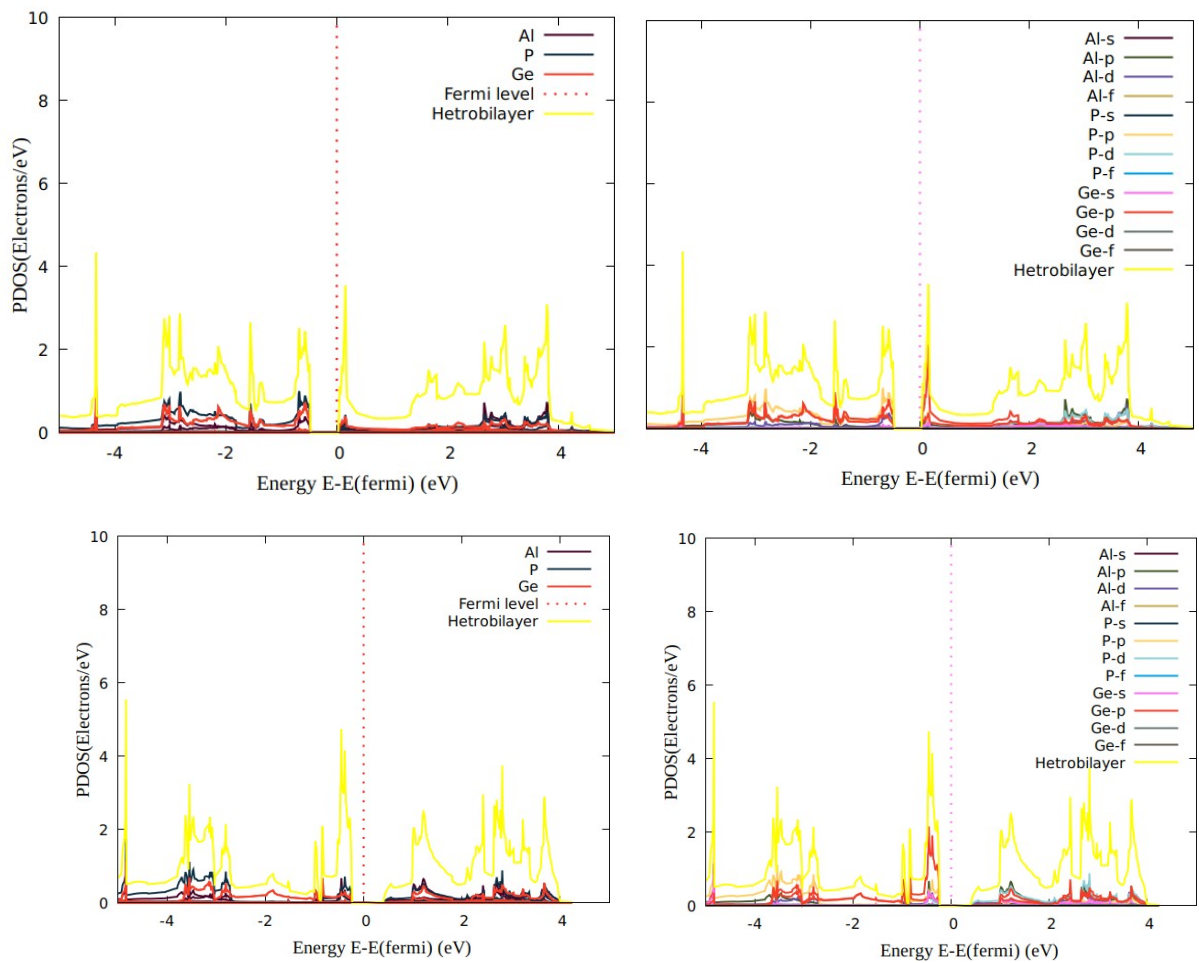

Figure 4.14: Atom projected density of states for structure-I and structure-II of the Ge/AlP heterobilayer. The Fermi level is made 0. Overall contribution from Al, P, and Ge atoms of (a) Structure-I and (b) Structure-II are exhibited.

The influence of Ge on the overall structure can further be demonstrated in Figure 4.15, which shows the charge density distribution for pattern-I. This clearly shows that the p-orbital of Ge is to be spread out, thus motivating the carrier concentration through the Ge layer only. Since AlP doesn't contribute much to carrier transportation, this also illustrates the expediency of choosing AlP as the substrate for the structure. Since all the structure resembles about the same distributions, we utilized pattern-II for the calculation of space charge density distribution and CDD.

From the charge density difference as shown in Figure 4.15 for pattern-I, the contribution of germanene at the space charge layer is clearly depicted. The charge contribution within the germanene layer changes, providing the breaking in symmetry. Moreover, electron transfers from the germanene layer to the AlP layer occur. This causes the evolve in a bandgap. Since most of the charge is convoluted in the bucked germanene structure, carrier mobility is controlled only by the germanene layer. This phenomenon observes well-matched with the work done during the investigation of C/SiC, Sn/SiC, C/MoS$_2$, C/h-BN 2D heterostructures [68,33]. Thus the figure gives insight into the resultant band gap and energy state.

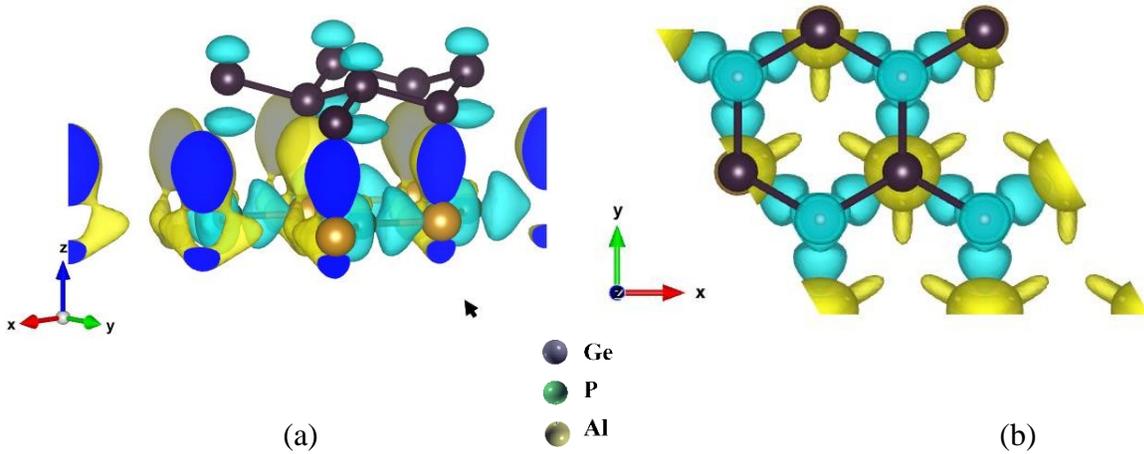

| | Ge |
| | P |
| | Al |

(a)                      (b)

Figure 4.15: The charge density difference of the Ge/AlP heterobilayer for Structure-I. Cyan and Yellow indicate the increase and decrease of charge in the space regarding free standing germanene and AlP layer, respectively. The isovalue is 0.004 e/Å3. (

Due to charge transfer within the germanene sublattices, the overall symmetry breaks down. Overlap of the layers causes the localizations of charges which may be the direct effect of orbital overlap or gradient of potential from layer on the close vicinity or may be due to polarization.

### 4.3.4 Band Gap Tuning with Variation of Interlayer Distance and Bi-Axial Strain

The germanene/AlP heterobilayer can be analyzed further for its bandgap tuning properties with respect to the interlayer distances, as demonstrated in Figure 4.16. With the increase in an interlayer distance, fewer interactions occur between the layers, less charge overlap & crystal symmetry distortion occurs. As a result of this phenomenon, less bandgap will induce at the Dirac cones. Again, if we reduce the interlayer distance from the convergence value, the band gap decreases, but this time, the band overlaps to a higher magnitude to compensate the tight-binding model and modern physics. The tenability of the bandgap in the heterostructure can be utilized for making spintronic devices quite efficiently as it is seen that the pattern-I & pattern-II follow similar characteristics throughout the investigation. The pattern-I with a slightly sluggish change in binding energy and band gap variation with interlayer distance in the range < 3.1 Å exhibits the change of the former two with a slightly lower amplitude. The bandgap tuning may also be achieved with a change in applied strain to the germanene layer, as shown in Figure 4.16. It is expected that the bandgap decreases slightly with an increase in strain associated with a significant fermi level, as observed by Asmaul *et al.* during the investigation of graphene/SiC heterostructure [68]. And an opposite change might be observed for compressive strain. More applied compressive or repulsive strain can cause the break in lattice symmetry and introduce variations in band gap at the Brillouin zone symmetrical points. From Table 4, it's clearly depicted that with an increase in interlayer distance, the bandgap is changeable, and a shift from indirect to direct bandgap at K-point at 1st BZ is observable. Below 3.5 Å, an indirect bandgap occurs between M point and a point between Gamma and K-point that shifts towards the K point with the increase in d. At 3.5 Å, indirect bandgap occurs between K and a point between Gamma and K – symmetrical points occur that tend towards K and finally results in a direct bandgap at K-point with further increase in 'd'.

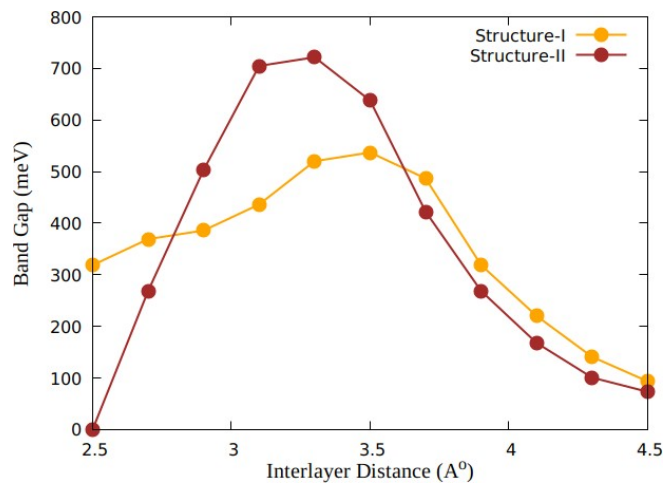

*Figure 4.16: Band tuning of Structure-I of Ge/AlP hterobilayer with variation in interlayer distance.*

Table 4.4: Level of bandgap, its position, and type with an increase in interlayer displacement d(Å) for Structure-I of Ge/AlP heterobilayer.

| Interlayer distance, d(Å) | Band Gap (meV) | Position of Band Gap (1st BZ) | Type of Band Gap |
|---|---|---|---|
| 2.5 | 318.8 | M+(G--K)) | indirect |
| 2.7 | 369.2 | M+(G--K) | indirect |
| 2.9 | 385.9 | M+(G--K) | indirect |
| 3.1 | 436.3 | M+(G--K) | indirect |
| 3.3 | 520.2 | M+(G--K) | indirect |
| 3.5 | 537.0 | K+(G--K) | indirect |
| 3.7 | 486.7 | K | direct |
| 3.9 | 318.8 | K | direct |
| 4.1 | 220.1 | K | direct |
| 4.3 | 140.8 | K | direct |
| 4.5 | 93.4 | K | direct |

# CHAPTER V

## Conclusion and Future Outlook

### 4.1 Conclusion

Two unique hetero-structures, germanene/2d-GaP, and germanene/2d-AlP, have been investigated in this research work. Their electronic properties are demonstrated by interpreting the outcomes obtained from DFT framework computations.

The germanene/2d-GaP heterostructure was proposed by showing three stacking patterns. It was found that this structure, when it becomes relaxed, possesses binding energy per germanene atom of about 63.9 meV, 215.9 meV, 90.02 meV, respectively. Binding energy larger than 180 meV suggests that rather than weak Van der Waals, the atoms of the bilayers are mutually connected by other strong bonds. The bandgap obtained with these is 463.2 meV, 285.1 meV, 154.4 meV. This bandgap being typically large in nature, provides their applicability in practice. The study on DOS, PDOS, and CDD shows that though due to the influence of GaP, germanene's lattice symmetry gets distorted in small magnitude and emerges bandgap, the properties of germanene is the one that will control the property of the heterostructure. The p-orbital of germanene is the most significant around the conduction band, showing that the carrier is only to move through the germanene. The charge is mostly accumulated on the germanene layer, exacerbating the utility of germanene. Incorporating SOC provides a significant change of the bandgap while maintaining the Dirac cone properties. In addition to that, the bandgap of the heterostructure can be tuned by changing interlayer distance and by applying biaxial strain. With the increase in the interlayer distance, there exists a shoot in the bandgap, which afterward reduces with a change in interlayer distance. Although there is no band gap whilst incorporating bi-axial compressive strain, with the increase in bi-axial tensile strain percentage, the bandgap reduces from the highest value and reaches 0 eV at 5%. These tuning properties excel the prediction of this heterostructure to be useful in the field of nano-electronics and spintronic devices.

As for the germanene/2d-AlP heterostructure, similar properties like that of germanene/2d-GaP result with a variation of the magnitude of binding energy and band gap levels, in case of this heterobilayer binding energy results about 102.5 meV, 190.9 meV, 129.7 meV, respectively with bandgap level 490 meV, 290.3 meV, 264.3 meV. In both cases, while tuning the bandgap, the bandgap transforms from indirect to direct with an increase in interlayer distance and becomes direct bandgap material at symmetrical Brillouin zone points, although with a lower bandgap. These properties of group-III phosphides and germanene bilayer may help in the growing nanomaterials revolution and will end up being the material of choice for semiconductor devices, sensors, LEDs, and many more.

## 4.2 Future Outlook

This research work was aimed to investigate the basic properties of the aforementioned two heterostructures. However, further research can be conducted in the following manner to wrap up all possible outcomes related to these structures:

- Planar configurations of germanene and 2d-GaP may be used to form a distinct heterobilayer, and investigation needs to perform on its properties.
- Germanene can be sandwiched between 2d-GaP or 2d-AlP or in between 2d-AlP and 2d- GaP layers & properties should be looked into.
- Emerging research interests like the Moire pattern of germanene can be taken into considerations with different rotational angles.
- To alter the property, impurities atoms and/or vacancies can be introduced to germanene layers.
- Applications of electric field on the germanene structure may be conducted. For HSE correlation and LDA, the electronics properties can be generated.

# SUPPLEMENTARY WORKS

1  First Principle Investigation of graphene/2d-GaN heterobilayer and study of its tunable electronic properties.
2  DFT observation of silicene/2d-GaAs heterostructure.
3  Study of planar silicene over planar GaP by Ab Initio method.
4  Tuning the electronic properties of wide bandgap materials: 2D-GaN over a 2D-SiC substrate.

# PUBLISHED WORKS

THE END